\documentclass{article}

\PassOptionsToPackage{numbers, sort&compress}{natbib}
\usepackage[main, final]{neurips_2026}
\usepackage[utf8]{inputenc} % allow utf-8 input
\usepackage[T1]{fontenc}    % use 8-bit T1 fonts
\usepackage{xcolor}         % colors
\definecolor{cusblue}{HTML}{000080}
\usepackage[colorlinks=true, % 启用颜色链接而不是方框
            linkcolor=cusblue,  % 引用链接颜色
            citecolor=cusblue
           ]{hyperref}
\usepackage{url}            % simple URL typesetting
\usepackage{booktabs}       % professional-quality tables
\usepackage{tabularx}
\usepackage{amsfonts}       % blackboard math symbols
\usepackage{nicefrac}       % compact symbols for 1/2, etc.
\usepackage{graphicx}
\usepackage{subfigure}
\usepackage{microtype}      % microtypography
\usepackage{bm}
\usepackage{bbm}
\usepackage{algorithm}
\usepackage{algorithmic}

\usepackage{wrapfig}

\usepackage{amsmath}
\usepackage{amssymb}
\usepackage{mathtools}
\usepackage{amsthm}

\theoremstyle{plain}
\newtheorem{theorem}{Theorem}[section]

\theoremstyle{definition}
\newtheorem{definition}[theorem]{Definition}
\newtheorem{assumption}[theorem]{Assumption}
\theoremstyle{remark}

\DeclareMathOperator*{\argmax}{arg\,max}

\title{LLM-Auction: Generative Auction towards LLM-Native Advertising}

\author{%
  Chujie Zhao, Qun Hu, Shiping Song, Dagui Chen\thanks{Corresponding author.}, Han Zhu, Jian Xu, Bo Zheng \\
  Taobao \& Tmall Group of Alibaba\\
  Beijing, China\\
  \texttt{zhaochujie.zcj@alibaba-inc.com, dagui.cdg@alibaba-inc.com} \\
}

\begin{document}

\maketitle

\begin{abstract}
The commercialization of LLM applications is the next frontier in online advertising, with \emph{LLM-native advertising} emerging as a promising paradigm by integrating ads into LLM-generated content. 
However, classic mechanisms are no longer applicable in this setting where the auction object is shifted from discrete ad slots to distributions over LLM outputs, and existing methods are impractical in industrial scenarios due to ignored externalities or high inference costs.
To address these issues, we propose LLM-Auction, the \emph{first} learning-based generative auction mechanism that \emph{integrates} auction and generation. 
By formulating the allocation as preference alignment between LLM outputs and a mechanism objective that balances advertiser value and user experience, we optimize the LLMs to inherently model allocation externalities without extra inference cost. 
Theoretically, we identify the allocation monotonicity and continuity of the optimal allocation rule of LLM-Auction, and prove that a simple first-price payment rule exhibits favorable incentive properties under this allocation rule. Furthermore, we build an LLM-as-a-judge simulation environment for quantitative evaluation, and experiments demonstrate that LLM-Auction achieves the state-of-the-art allocation efficiency while approximately satisfying key mechanism properties.
\end{abstract}

\section{Introduction}
\label{sec:intro}

The rise of chatbots and AI search engines powered by large language models (LLMs)~\citep{llm_survey, gpt3} is fundamentally reshaping how users acquire information~\citep{chatgpt_replace_search}. As these applications become primary gateways for user traffic, the development of compatible commercialization strategies has become increasingly critical.
Current industrial efforts adopt \emph{external} ad placements, such as fixed ad slots attached to LLM responses~\citep{perplexity,microsoft_gen_ad,google_sge}. These formats are easy to deploy, but do not take advantage of the capabilities of LLMs in reasoning and generation.

We focus on \emph{LLM-native advertising}, where promotional content is naturally integrated into an LLM response while balancing advertiser value and user experience.
This setting changes the allocation object from discrete slots to a distribution over LLM outputs: advertisers bid to influence this distribution, increasing the likelihood of their desired content appearing in the generated response.
This shift renders classic mechanisms like Generalized Second-Price (GSP) auction~\citep{gsp} inapplicable, creating urgent demand for tailored auction mechanisms.

\begin{figure}[t]
  \centering
    \includegraphics[width=1.0\textwidth]{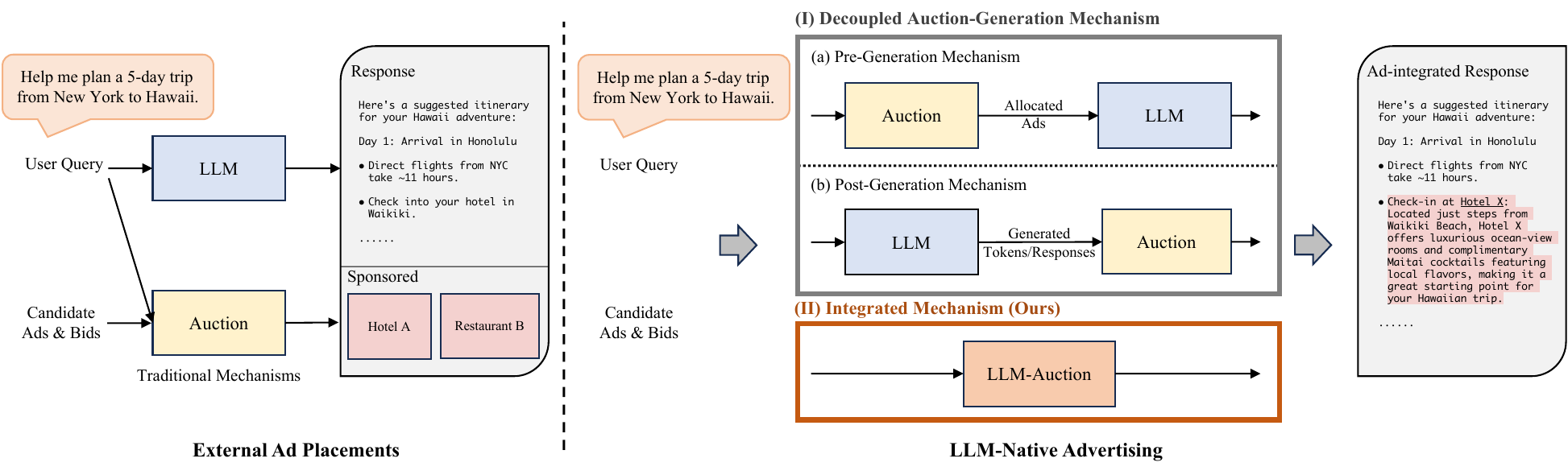}
    \caption{
      Comparison of ad formats and auction mechanisms in LLM-based AI applications.
    }
    \label{fig:intro}
\end{figure}

Existing mechanisms for LLM-native advertising mostly decouple auction from generation~\citep{token_auction, MOSAIC, rag_auction, summary_auction}. As shown in Figure~\ref{fig:intro}, \emph{pre-generation} mechanisms choose ads before generation, requiring preset ad counts or insertion structures and weakening their ability to model response-level externalities, leading to suboptimal performance.
\emph{Post-generation} mechanisms generate ad-conditioned tokens or candidate responses before running an auction. They are more context-aware, but require extra LLM forward passes whose cost grows with the number of advertisers or candidates, limiting practicality in high-concurrency serving systems.

To address these challenges, we propose LLM-Auction, the \emph{first} learning-based generative auction mechanism that \emph{integrates} auction and LLM generation for LLM-native advertising. Its key idea is to make the LLM itself the allocation rule.
Conditioned on the user query, profile, candidate ads, and bids, LLM-Auction post-trains the model to prefer responses that maximize a mechanism objective balancing advertiser value and user experience.
We instantiate the objective with a reward model based on advertising metrics such as click-through rate, and introduce \emph{Iterative Reward-Preference Optimization} (IRPO), which alternates between reward model updates from feedback and LLM updates from reward-induced preferences.
At serving time, one LLM forward pass generates the ad-integrated response, internalizing response-level externalities without extra inference.

We further show that the optimal LLM-Auction allocation satisfies monotonicity and continuity, linking response-distribution allocation to advertiser incentives.
With a simple first-price payment rule, which avoids the costly computation of the dominant-strategy incentive-compatible (DSIC) payment, these properties guarantee a mixed-strategy equilibrium with an explicit bid-shading condition for utility maximizers and effective-value incentive compatibility for ROI-constrained value maximizers under the optimal allocation rule.

Finally, inspired by the LLM-as-a-judge paradigm~\citep{llm_as_a_judge_survey}, we build a simulation environment covering user-query generation, ad-integrated response generation, and user feedback for large-scale training and quantitative evaluation.
Extensive experiments demonstrate that LLM-Auction \emph{significantly} outperforms existing mechanisms in allocation efficiency while approximately satisfying key mechanism properties.

Our main contributions are summarized as follows:
\begin{itemize}
    \item We lay down general definitions and incentive desiderata for LLM-native advertising auction mechanisms, establishing foundations for future work.
    \item We propose LLM-Auction, the \emph{first} learning-based generative auction mechanism that integrates allocation and generation through preference alignment.
    \item We identify the allocation monotonicity and continuity of the optimal allocation rule of LLM-Auction, and prove that a simple first-price payment rule exhibits favorable incentive properties under this allocation rule.
    \item We develop an LLM-as-a-judge simulation environment for scalable data construction and quantitative evaluation of LLM-native advertising mechanisms.
\end{itemize}

\section{Related Work}
\label{sec:related}

\noindent\textbf{Position auctions in online advertising.}  
\citet{position_auction} established the game-theoretic foundation of position auctions in sponsored search, and \citet{gsp} formalized the GSP auction with equilibrium and efficiency guarantees. \citet{caragiannis_gsp_efficiency} further prove that GSP achieves a constant-factor approximation to optimal social welfare even under worst-case equilibria.
Recent work extends GSP with learning-based mechanisms~\citep{deepgsp,dna}, while \citet{cga} propose a multi-slot generative auction that accounts for externalities.

\noindent\textbf{Mechanism design for LLMs.} 
Mechanism design for LLMs studies auction mechanisms for LLM-generated content with desirable incentive properties. Existing methods typically decouple auction from generation, falling into two categories.
(i) \textit{Pre-generation mechanisms} allocate ads before generation: \citet{summary_auction} introduce a prominence-based auction for LLM summarization, \citet{rag_auction} design RAG-based mechanisms, and \citet{position_auction_in_genai} extend position auctions to AI-generated content.
These mechanisms must specify ad counts or positions upfront and do not model allocation externalities, such as context and co-exposure effects.
(ii) \textit{Post-generation mechanisms} first produce ad-conditioned candidates, then select or aggregate them by auction. \citet{feizi2023online} and \citet{sqa} add ad rewriting to position auctions; \citet{token_auction} aggregate advertisers' preferences during decoding; and \citet{MOSAIC} select among response-level candidates.
Although more context-aware, these methods require extra LLM inference that scales with the number of advertisers or candidate responses, limiting high-concurrency deployment.
\citet{data_driven_vcg} also adapt VCG mechanisms to LLM settings from a theoretical perspective.

\noindent\textbf{LLM alignment methods.}  
LLM alignment steers model outputs toward external objectives or human preferences. RLHF~\citep{rlhf_paper1,rlhf_2} optimizes LLMs with learned preference reward models, while DPO~\citep{dpo} derives a closed-form objective for direct optimization from preference pairs. Subsequent variants, including IPO~\citep{ipo}, KTO~\citep{kto}, and SimPO~\citep{simpo}, refine this objective, while tuning-free methods such as LA~\citep{linear_alignment} and Amulet~\citep{zhang2025amulet} adapt preferences at inference time through latent representations.

\section{Preliminaries}
\label{pre}

We formalize an LLM-native advertising auction as a single-turn interaction between a user and an LLM. Let $[n]$ denote the advertisers. Each advertiser $i$ has a private value $v_i \in \mathbb{R}_{\geq 0}$ and a fixed ad $a_i \in \mathcal{A}$; write $\bm{v}=(v_1,\ldots,v_n)$ and $\bm{a}=(a_1,\ldots,a_n)$. The context is $s=(x,h,\bm{a}) \in \mathcal{S}$, where $x \in \mathcal{X}$ is the user query and $h \in \mathcal{H}$ is the user profile.

Given context $s$ and bid profile $\bm{b}=(b_1,\ldots,b_n) \in \mathbb{R}_{\geq 0}^n$, the LLM $\pi_\theta$ defines an allocation distribution $\pi_\theta(\cdot \mid s,\bm{b}) \in \Delta(\mathcal{Y})$ over responses. A realized response $y \sim \pi_\theta(\cdot \mid s,\bm{b})$ may contain content related to any subset of candidate ads.
An auction mechanism is a tuple $\mathcal{M}=(\pi_\theta,\bm{p})$:
\begin{itemize}
    \item \textbf{Allocation rule}: $\pi_\theta: \mathcal{S} \times \mathbb{R}_{\geq 0}^n \to \Delta(\mathcal{Y})$ maps the context and bids to a response distribution.
    \item \textbf{Payment rule}: $\bm{p}: \Delta(\mathcal{Y}) \times \mathbb{R}_{\geq 0}^n \to \mathbb{R}_{\geq 0}^n$ maps an allocation distribution and bids to payments $\bm{p}(\pi_\theta(\cdot \mid s,\bm{b}),\bm{b})$.
\end{itemize}
To measure the effectiveness of ad $a_i$ in response $y$ and context $s$, let $f_i(y,s)$ denote a performance metric such as click-through rate (CTR) or conversion rate (CVR). We extend this notation to distributions by writing $f_i(\pi_\theta(\cdot \mid s,\bm{b}),s):=\mathbb{E}_{y \sim \pi_\theta(\cdot \mid s,\bm{b})}[f_i(y,s)]$. A higher bid should not reduce an advertiser's expected performance, yielding the following desideratum.
\begin{definition}[Allocation Monotonicity]
\label{def:allocation_monotonicity}
An allocation rule $\pi_\theta$ is \emph{monotone} if, for every advertiser $i$, context $s$, competing bids $\bm{b}_{-i}$, and bids $b_i' \geq b_i$,	
\[
f_i(\pi_\theta(\cdot \mid s,(b_i', \bm{b}_{-i})),s) \geq f_i(\pi_\theta(\cdot \mid s,(b_i, \bm{b}_{-i})),s).
\]
\end{definition}
Advertisers optimize bids under different objectives and constraints. We consider two standard models: utility maximizers (UM) and value maximizers (VM).
\begin{definition}[Utility Maximizer (UM)]
\label{def:UM}
A utility-maximizing advertiser $i$ chooses $b_i \geq 0$ to maximize expected quasi-linear utility:
\begin{equation}
u_i(b_i; s, v_i, \bm{b}_{-i}) = 
v_i \cdot f_i(\pi_\theta(\cdot \mid s,(b_i,\bm{b}_{-i})),s) 
-  p_i(\pi_\theta(\cdot \mid s,(b_i,\bm{b}_{-i})),(b_i,\bm{b}_{-i})).
\end{equation}
\end{definition}
\begin{definition}[Value Maximizer (VM)]
\label{def:VM}
Let $\tau_i > 0$ be advertiser $i$'s target ROI. A value maximizer chooses $b_i \geq 0$ to maximize expected value subject to the ROI constraint:
\begin{equation}
\begin{aligned}
\max_{b_i \geq 0} \quad &v_i \cdot f_i(\pi_\theta(\cdot \mid s,(b_i,\bm{b}_{-i})),s) \\
\text{s.t.} \quad &\tau_i \cdot p_i(\pi_\theta(\cdot \mid s,(b_i,\bm{b}_{-i})),(b_i,\bm{b}_{-i})) \leq v_i \cdot f_i(\pi_\theta(\cdot \mid s,(b_i,\bm{b}_{-i})),s).
\end{aligned}
\end{equation}
% For notation consistency, we also refer to the VM objective as ``utility'' $u_i$.
\end{definition}
The following properties capture incentives and participation.
\begin{definition}[Incentive Compatibility for Utility Maximizers (UM-IC)]
\label{def:IC}
A mechanism is \emph{UM-IC} if truthful value reporting is a dominant strategy: for every $i$, $s$, $\bm{b}_{-i}$, and $b_i \geq 0$, $u_i(v_i;s,v_i,\bm{b}_{-i}) \geq u_i(b_i;s,v_i,\bm{b}_{-i})$.
\end{definition}
\begin{definition}[Effective-Value Incentive Compatibility for Value Maximizers (VM-IC)]
\label{def:vm_ic}
For a value maximizer with type $(v_i,\tau_i)$, define the effective value $\tilde{v}_i=v_i/\tau_i$. A mechanism is \emph{VM-IC} if, for every $i$, $s$, $\bm{b}_{-i}$, $b_i=\tilde{v}_i$ is a dominant strategy for maximizing expected value subject to the ROI constraint. Thus, truthful reporting in VM means reporting effective value rather than raw value $v_i$.
\end{definition}
\begin{definition}[Individual Rationality (IR)]
\label{def:IR}
A mechanism is \emph{individually rational} if, for every advertiser, context, and bid profile, expected payment is at most expected value:
\[
p_i(\pi_\theta(\cdot \mid s,\bm{b}),\bm{b}) \leq v_i \cdot f_i(\pi_\theta(\cdot \mid s,\bm{b}),s).
\]
\end{definition}

The designer seeks an allocation policy that balances advertiser value with user experience:
\begin{equation}
{\small
    \resizebox{0.94\linewidth}{!}{$\displaystyle J(\pi_\theta,\bm{v},\bm{b},s)=\sum_{i=1}^n v_i \cdot f_i(\pi_\theta(\cdot \mid s,\bm{b}),s)+\lambda \cdot \Omega_{\text{resp}}(\pi_\theta(\cdot \mid s,\bm{b}),s)-\beta \cdot \mathcal{D}_{\text{KL}}(\pi_\theta(\cdot \mid s,\bm{b}) \| \pi_{\theta_0}(\cdot \mid s,\bm{b})).$}
}
    \label{eq:objective}
\end{equation}
Here, $\Omega_{\text{resp}}(\pi_\theta(\cdot \mid s,\bm{b}),s)$ is the expected penalty for ad-insertion constraint violations, and $\mathcal{D}_{\text{KL}}$ penalizes deviations from the reference LLM $\pi_{\theta_0}$, i.e., the pre-trained LLM before any auction-oriented training, to preserve its response quality.

\begin{definition}[Optimal LLM-Native Advertising Mechanism]
In the LLM-native advertising setting, the optimal mechanism $\mathcal{M}^*=(\pi_{\theta^*},\bm{p}^*)$ consists of: (i) an allocation rule $\pi_{\theta^*}$ that maximizes Eq.~\eqref{eq:objective} for every realization $(\bm{v},\bm{b},s)$;
and (ii) a payment rule $\bm{p}^*$ that ensures the relevant IC and IR.
\label{def:optimal_mechanism}
\end{definition}

Since values $\bm{v}$ are private and unobservable, the mechanism uses the reported bid vector $\bm{b}$ as the value signal, yielding the bid-based objective $J(\pi_\theta,\bm{b},\bm{b},s)$.
In industrial deployment, this optimal mechanism may be relaxed for computational tractability; for example, strict IC may be relaxed to support an efficient payment rule.

\section{LLM-Auction}
\label{sec:auction_gpt}

We now introduce LLM-Auction, the first integrated generative auction mechanism for LLM-native advertising. Full proofs are deferred to Appendix~\ref{appendix:proofs}, and Appendix~\ref{appendix:assumption_feasibility} discusses the plausibility of the assumptions underlying the theoretical results.

\subsection{Allocation Rule}
\label{subsec:allocation}

Section~\ref{pre} defines the allocation result as the LLM response distribution. LLM-Auction parameterizes the allocation rule directly with $\pi_\theta$, so allocation optimization amounts to shifting $\pi_\theta(\cdot \mid s,\bm{b})$ toward the objective in Eq.~\eqref{eq:objective}. 
% Since private values are not observed by the mechanism, LLM-Auction uses the reported bid vector $\bm{b}$ as the value signal when optimizing allocations.
We cast this as preference alignment: the LLM is fine-tuned on contexts, candidate ads, and bids using a reward model aligned with the mechanism objective.

\subsubsection{Mechanism Reward Model}
\label{subsubsec:reward_model}

Preference alignment requires comparing alternative ad-integrated responses for the same input $(s,\bm{b})$, but user logs reveal feedback only for realized responses. We therefore build a reward model from a performance metric predictor, which estimates the advertiser value of candidate responses during offline LLM training.

\noindent \textbf{Performance metric prediction.} We use a predictor $\hat{f}^\psi$ to estimate response-level advertising performance. Unlike serving-time prediction models, $\hat{f}^\psi$ is used only for offline LLM training and can condition on the generated response $y$. Let $\hat{f}^{\psi}_i(y,s)$ denote the predicted metric for ad $i$ under response-context pair $(y,s)$. We train it on feedback logs with binary cross-entropy:
\begin{equation}
    \mathcal{L}_{\text{BCE}} = -\sum_{i=1}^n \left( c_i \log\hat{f}_i^{\psi}(y,s)+ (1 - c_i) \log(1 - \hat{f}_{i}^{\psi}(y,s)) \right),    
    \label{eq:bce_loss}
\end{equation}
where $c_i \in \{0,1\}$ is the feedback label for ad $i$ (e.g., click or conversion). To connect the learned reward to the mechanism objective, we use the following idealization:
\begin{assumption}[Unbiased performance prediction]
\label{def:unbiased pctr}
    The prediction model is unbiased under any allocation rule $\pi_\theta$ if, for any advertiser $i$, context $s$, and bid vector $\bm{b}$,
    \[
    \mathbb{E}_{y \sim \pi_\theta(\cdot \mid s, \bm{b})}[\hat{f}_i^{\psi}(y, s)] = \mathbb{E}_{y \sim \pi_\theta(\cdot \mid s, \bm{b})}[f_i(y, s)]
    \]
\end{assumption}

\noindent\textbf{Reward model.} 
We fine-tune the LLM with Direct Preference Optimization (DPO)~\citep{dpo} because it offers a stable preference-learning objective with a KL-regularized reward-maximization interpretation. Taking bids as the value signal available to the mechanism and using Assumption~\ref{def:unbiased pctr}, we define the response-level reward:
\begin{equation}
{\small
    \mathcal{R}(s, \bm{b}, y) 
    = \sum_{i=1}^n b_i \cdot \hat{f}^{\psi}_i(y, s)
    + \lambda \cdot \Omega_{\text{resp}}(y),
    \label{eq:reward_model}
}
\end{equation}
where $\Omega_{\text{resp}}(y)$ penalizes ad-insertion constraint violations in the generated response. This reward provides preference feedback for aligning the LLM with the desired allocation rule.

\subsubsection{Iterative Reward-Preference Optimization Algorithm}
\label{subsubsec:iterative_algo}

\begin{figure}[ht]
  \centerline{\includegraphics[width=0.7\textwidth]{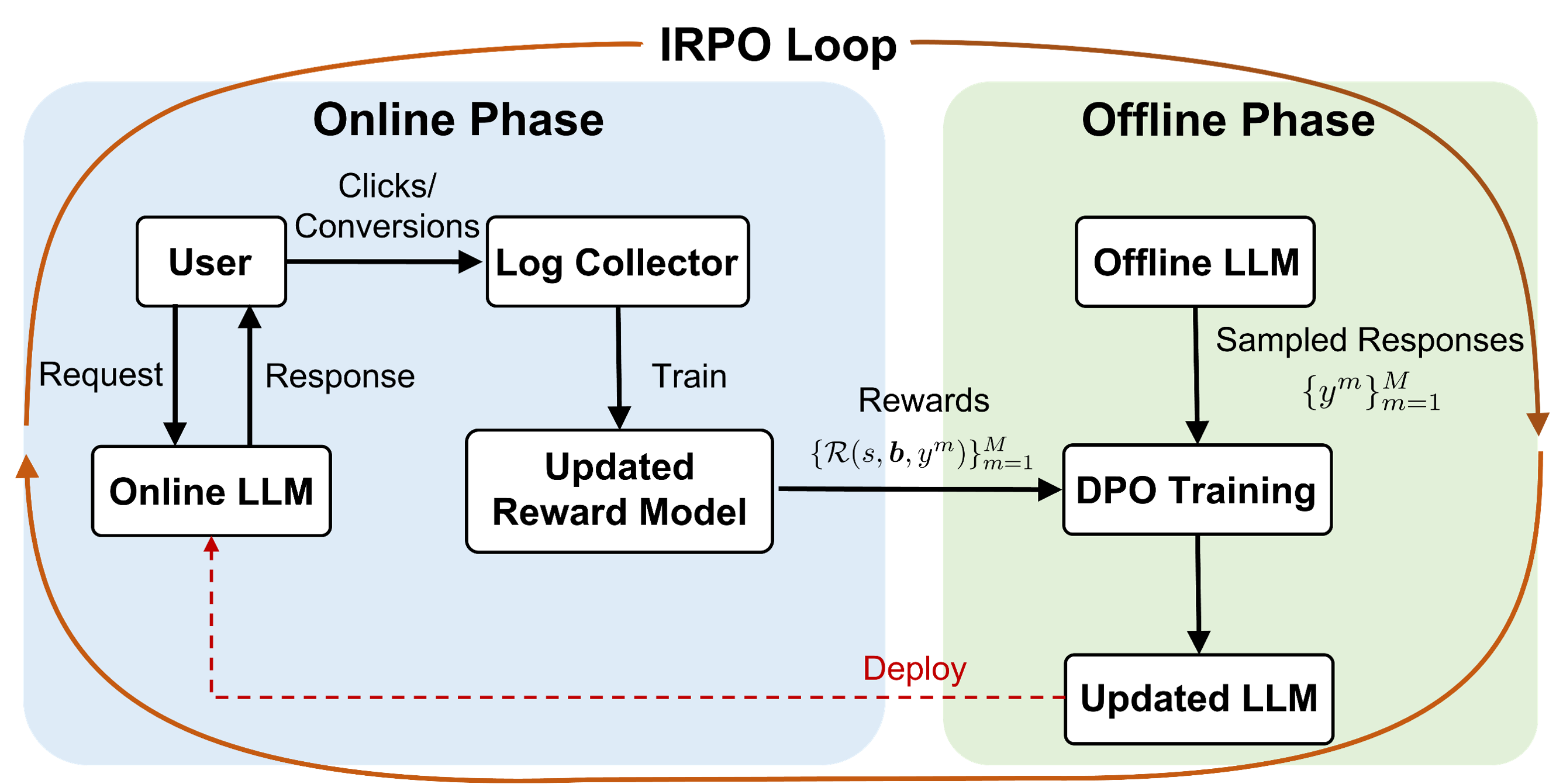}}
  \caption{Illustration of the iterative optimization framework of LLM-Auction.}
  \label{fig:irpo_method}
\end{figure}

Direct optimization of Eq.~\eqref{eq:reward_model} with a fixed predictor can suffer from a distribution shift: as the LLM policy changes, its responses move away from the data used to train $\hat{f}^\psi$. We address this with Iterative Reward-Preference Optimization (IRPO), shown in Figure~\ref{fig:irpo_method}, which alternates between updating the reward model from user feedback and updating the LLM with reward-induced preferences.

Let $\pi_{\theta_t}$ and $\mathcal{R}_{\psi_t}$ denote the LLM and reward model at epoch $t$. IRPO has two phases:

\noindent\textbf{Reward model update.} We freeze $\pi_{\theta_{t-1}}$, deploy it to collect online user feedback, and construct $\mathcal{D}_{\text{on}}^t=\{s^k,y^k,\bm{c}^k\}_{k=1}^{N_{\text{on}}}$. We then optimize the predictor with Eq.~\eqref{eq:bce_loss}, yielding the updated reward model $\mathcal{R}_{\psi_t}$.

\noindent\textbf{LLM update.} 
We fix $\mathcal{R}_{\psi_t}$ and construct $\mathcal{D}_{\text{off}}^t$ from $\mathcal{D}_{\text{on}}^t$ or synthetic data. For each $(s,\bm{b}) \in \mathcal{D}_{\text{off}}^t$, we sample $M$ responses $y^m \sim \pi_{\theta_{t-1}}(\cdot \mid s,\bm{b})$ and score each response with $r^m=\mathcal{R}_{\psi_t}(s,\bm{b},y^m)$.
The highest-reward response $y^w$ is the winner, and responses with a reward gap above threshold $\delta_{\text{th}}$ form the losing set $\mathcal{K}=\{y^l \mid r^w-r^l>\delta_{\text{th}}\}$. The DPO loss is:
\begin{equation}
{\small
\mathcal{L}_{\text{DPO}} = - \frac{1}{|\mathcal{K}|} \sum_{y^l \in \mathcal{K}} \Bigg[ \log \sigma\Bigg(   \beta\log \frac{\pi_{\theta_{t}}(y^w|s,\bm{b})}{\pi_{\theta_{t-1}}(y^w|s,\bm{b})} 
 - \beta \log \frac{\pi_{\theta_{t}}(y^l|s,\bm{b})}{\pi_{\theta_{t-1}}(y^l|s,\bm{b})} \Bigg) \Bigg],
}
\label{eq:dpo_loss_irpo}
\end{equation}
where $\beta$ controls the KL penalty. Minimizing Eq.~\eqref{eq:dpo_loss_irpo} gives the next LLM $\pi_{\theta_t}$.

IRPO progressively refreshes reward estimation under the current allocation policy, reducing distribution shifts and improving allocation performance. Algorithm~\ref{alg:irpo} provides the pseudocode.

\subsubsection{Properties of Allocation Rule}
\label{subsubsec:alloc-prop}

Next, we analyze the theoretical properties of the optimal allocation rule $\pi_{\theta^*}$ induced by Eq.~\eqref{eq:objective}. Throughout the theoretical analysis, we impose the following assumption on the reference LLM $\pi_{\theta_0}$.
\begin{assumption}[Bid Independence of the Reference LLM Distribution]
\label{assump:bid_independence}
For every context $s$, response $y$, and bid profile
$\bm b \in [0,\bar b]^n$, the reference LLM distribution satisfies
\[
\pi_{\theta_0}(y \mid s,\bm b)=\pi_{\theta_0}(y \mid s).
\]
\end{assumption}
Intuitively, the reference LLM has not yet learned how bids should shape its output distribution. The experimental results in Figure~\ref{fig:property}(a) also provide empirical support for this assumption.

\begin{theorem}[Monotonicity of the Optimal Allocation]
\label{thm:alloc-mono}
The optimal allocation $\pi_{\theta^*}$ that maximizes the bid-based mechanism objective $J(\pi_\theta,\bm{b},\bm{b},s)$ satisfies allocation monotonicity. Formally, for any advertiser $i$, fixed context $s$, and competing bids $\bm{b}_{-i}$, if $b_i' \geq b_i$, then:
\[
f_i(\pi_{\theta^*}(\cdot \mid s, b_i', \bm{b}_{-i}), s) \geq f_i(\pi_{\theta^*}(\cdot \mid s, b_i, \bm{b}_{-i}), s).
\]
\end{theorem}

Theorem~\ref{thm:alloc-mono} states that increasing a bid cannot reduce the advertiser's expected performance, providing the allocation-side foundation for incentive analysis.

\begin{theorem}[Continuity of the Optimal Allocation]
\label{thm:continuity}
In the LLM-native advertising setting, under Assumption~\ref{assump:bid_independence}, the expected performance of the optimal allocation rule $\pi_{\theta^*}$ satisfies continuity: $f_i(\pi_{\theta^*}(\cdot \mid s, \bm{b}), s)$ is continuous in $\bm{b}$ for all $i, s$.
\end{theorem}

We will discuss the practical plausibility of Assumption~\ref{assump:bid_independence} in Appendix~\ref{appendix:assumption_feasibility} and empirically validate allocation continuity in Section~\ref{subsubsec:alloc_continuity}.

\subsection{First-Price Payment Rule}
\label{subsec:payment}

For utility maximizers (UMs), the monotonicity of the optimal allocation rule $\pi_{\theta^*}$ admits a DSIC payment through Myerson's payment identity~\cite{myerson1981optimal}. Let $A_i(\bm b;s):=f_i(\pi_{\theta^*}(\cdot\mid s,\bm b),s)$ denote advertiser $i$'s effective allocation. With the normalization $P_i^{\mathrm{DSIC}}(0,\bm b_{-i};s)=0$, the corresponding expected payment is
\[
P_i^{\mathrm{DSIC}}(\bm b;s)=b_i A_i(\bm b;s)-\int_0^{b_i} A_i(z,\bm b_{-i};s)\,dz.
\]
By the Gibbs form of $\pi_{\theta^*}$, this integral equals $\beta$ times the log-ratio of two partition functions (Appendix~\ref{appendix:dsic}). However, exactly evaluating these partition functions over the LLM response space is computationally expensive.

We therefore adopt the operationally simpler first-price payment rule:
\begin{equation}
p_i(\pi_\theta(\cdot\mid s,\bm b),\bm b)=b_i \cdot f_i(\pi_\theta(\cdot\mid s,\bm b),s).
\end{equation}
Given the allocation, this payment requires only one multiplication per advertiser, avoiding the bid integral and counterfactual partition functions of the DSIC payment. Although first-price payments are generally not DSIC for UMs, the bidding game they induce under the optimal allocation rule admits a mixed-strategy Nash equilibrium, whose interior bids satisfy a bid-shading condition.

\begin{theorem}[UM Equilibrium under First-Price Payment Rule]
\label{thm:um_equilibrium}
Under the optimal allocation rule $\pi_{\theta^*}$ with compact bid spaces, the first-price bidding game among UMs admits a mixed-strategy Nash equilibrium.
At any such equilibrium $\sigma^\star$, every interior support bid $b_i$ with $\overline V_i(b_i)>0$ satisfies
\begin{equation}
b_i = v_i-\beta \frac{\overline A_i(b_i)}{\overline V_i(b_i)}.
\label{eq:um_bid_shading}
\end{equation}
\end{theorem}

Here,
$\overline A_i(b_i):=\mathbb{E}_{\bm b_{-i}\sim\sigma_{-i}^\star}[A_i(b_i,\bm b_{-i};s)]$
and
$\overline V_i(b_i):=\mathbb{E}_{\bm b_{-i}\sim\sigma_{-i}^\star}[\operatorname{Var}_{y\sim\pi_{\theta^*}(\cdot\mid s,b_i,\bm b_{-i})}[f_i(y,s)]]$
are advertiser $i$'s expected allocation and expected performance variance when it bids $b_i$ against $\sigma_{-i}^\star$.

The intuition behind Eq.~\eqref{eq:um_bid_shading} is a marginal cost--benefit balance. Raising $b_i$ increases the payment on advertiser $i$'s existing expected allocation (marginal cost $\overline A_i(b_i)$), but also tilts $\pi_{\theta^*}$ toward responses with above-average $f_i(y,s)$, buying additional allocation at rate $\overline V_i(b_i)/\beta$, each unit worth the net surplus $v_i-b_i$ (marginal benefit $(v_i-b_i)\overline V_i(b_i)/\beta$). An interior support bid balances the two exactly, so advertiser $i$ shades its bid below its value $v_i$ by $\beta\overline A_i(b_i)/\overline V_i(b_i)$, and shades more when its allocation is less responsive to its bid (smaller $\overline V_i(b_i)$) or when its existing allocation is already large.

For value maximizers (VMs), we prove that truthful reporting of effective value is an optimal strategy under the first-price payment rule:
\begin{theorem}[VM-IC under First-Price Payment Rule and an ROI Constraint]
\label{thm:vm_roi}
In the LLM-native advertising setting with the optimal allocation rule, the first-price payment rule is VM-IC under ROI constraint $\tau_i$, where the optimal bidding strategy is $b_i=\tilde{v}_i=v_i /\tau_i$.
\end{theorem}
Furthermore, individual rationality (IR) is naturally satisfied under standard no-overbidding behavior for UMs and when $\tau_i \ge 1$ for VMs under the first-price payment rule.

In summary, the first-price payment rule is computationally efficient, satisfies IR, admits equilibria for UMs, and achieves VM-IC for ROI-constrained VMs.

\section{Experiments}
\label{sec:experiments}

\setcounter{footnote}{0}
\renewcommand{\thefootnote}{\arabic{footnote}}

\subsection{LLM-Native Advertising Simulation Environment}
\label{subsec:benchmark}

A key bottleneck in mechanism design for LLM-native advertising is the absence of standardized, quantitative evaluation at scale. Public logs for ad-integrated LLM responses are not yet available, while prior studies often rely on qualitative examples or small-scale tests that do not reflect the scale and complexity of real-world industrial ad auctions.

\begin{figure}[ht]
  \begin{center}
    \centerline{\includegraphics[width=0.9\linewidth]{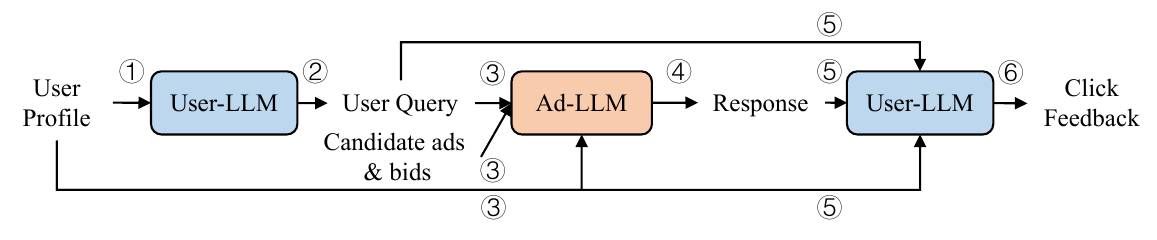}}
    \caption{Simulation environment for LLM-native advertising. The User-LLM generates personalized queries and click feedback, while the Ad-LLM produces ad-integrated responses under the evaluated mechanism. Numbers indicate data flow.}
    \label{fig:bench}
  \end{center}
  % \vspace{-2.5em}
\end{figure}

To bridge this gap, we build a simulation environment, inspired by the LLM-as-a-Judge paradigm~\citep{llm_as_a_judge_survey}, that covers the full ad-serving and feedback pipeline for LLM applications such as chatbots and AI search. As shown in Figure~\ref{fig:bench}, our pipeline consists of two modules: First, the Ad-LLM generates ad-integrated responses under the evaluated mechanism; for LLM-Auction, this Ad-LLM is the trained allocation model itself. Second, the User-LLM generates personalized queries and returns click feedback conditioned on the user profile, query, and generated response.

The pipeline involves three stages: (i) \textbf{User query generation:} the User-LLM generates user queries based on user profiles containing demographic information and interests. (ii) \textbf{Ad-integrated response generation:} conditioned on the user query, profile, and \emph{all} candidate ads with bids, the Ad-LLM generates a response that integrates clickable ads according to the mechanism.
\footnote{To highlight the ad and facilitate its recognition by the User-LLM, we adopt the insertion format \texttt{@ad\_title@[ad\_id]}. Currently, we restrict the Ad-LLM from modifying ad titles to ensure faithfulness, leaving unconstrained advertising to future work.}
(iii) \textbf{User click feedback:} the User-LLM simulates clicks and returns structured feedback for every inserted ad. Appendix~\ref{appendix:env} provides an example, prompt templates, and configuration details of the environment.

User-LLM feedback inevitably reflects a biased simulator distribution. We therefore view the environment as a controlled sandbox, not a substitute for real user feedback. This limitation does not couple LLM-Auction to the simulator: the method only requires click-based feedback and can optimize against any supplied click distribution. Once online logs from real users become available, the same pipeline can train directly on those authentic feedback signals. A small-scale human validation shows that the User-LLM's click decisions agree with human judgments well above chance (Appendix~\ref{appendix:human_validation}), and we empirically verify the robustness of LLM-Auction against different user feedback distributions in Appendix~\ref{appendix:sim_robustness}.

\subsection{Experimental Setup}
\subsubsection{Dataset}
We construct 3k synthetic user profiles with demographic information and interests. The User-LLM generates five queries per profile, yielding 15k queries. We keep this query set fixed across experiments and split it into 12k training queries and 3k test queries. The candidate ad set contains 100 real-world advertisements from a leading e-commerce platform, spanning 50 categories.

\subsubsection{Baselines}
Following the protocol used by learning-based mechanisms~\citep{deepgsp,dna,swvcg,cga}, we evaluate all methods on fixed single-round bids. We compare LLM-Auction with the pre-trained LLM, RAG-Auction~\citep{rag_auction}, MOSAIC~\citep{MOSAIC}, and LLM-Auction (Oracle), an upper-bound variant using direct feedback. See Appendix~\ref{appendix:baseline} for detailed evaluation protocol and implementation details of all baselines.

\subsubsection{Implementation Details of LLM-Auction}
\label{subsubsec:imp_detail}

\noindent\textbf{Reward design.}
Following Eq.~\eqref{eq:reward_model}, we instantiate $\hat{f}_i^{\psi}$ as a predicted click-through rate (pCTR) model, while $\Omega_{\text{resp}}(y)$ penalizes excessive ad insertion and formatting errors. This lightweight reward isolates allocation efficiency, but the same interface can incorporate ad fidelity, credibility, hallucination, bias, or trust signals. Appendix~\ref{appendix:exp_details} details the pCTR and reward models.

\noindent\textbf{Training protocol.}
We train Qwen3-4B-Instruct~\citep{yang2025qwen3} with IRPO. During each reward-model update stage, we use the previous epoch's Ad-LLM to sample 10 responses per query, and collect the User-LLM feedback to form $\mathcal{D}_{\text{on}}$ for the pCTR model updates. During each LLM update stage, we sample 5 responses per training query, score them with the updated reward model, and optimize the Ad-LLM. Each prompt of the Ad-LLM contains the full candidate ad set, and bids are resampled uniformly from positive integers in $[1,100]$ at every epoch to avoid overfitting to static bid patterns. Appendix~\ref{appendix:detail_llm_auction} provides additional implementation details and hyperparameters.

\subsubsection{Metrics}
We evaluate mechanisms on held-out test queries $\mathcal{D}_{\text{test}}$ with independently sampled bid profiles. Since click events trigger realized payments in our simulator, let $z_i^k$, $c_i^k$, $b_i^k$, and $p_i^k$ denote the impression indicator, click indicator, bid, and payment for advertiser $i$ on sample $k$, respectively, with $p_i^k=b_i^k \cdot c_i^k$. We compute the following allocation efficiency metrics:
\begin{itemize}
    \item Revenue per Query: $\frac{1}{|\mathcal{D}_{\text{test}}|} \sum_{k \in \mathcal{D}_{\text{test}}} \sum_{i=1}^n p_i^k$.
    \item Reward per Query: $\frac{1}{|\mathcal{D}_{\text{test}}|} \sum_{k \in \mathcal{D}_{\text{test}}} \mathcal{R}(s^k,\bm{b}^k,y^k)$.
\end{itemize}

\subsection{Comparison of Allocation Efficiency}
\label{subsec:efficiency_comparison}

\begin{table}[ht]
\caption{Comparison of allocation efficiency in the LLM-native advertising simulation environment. Values are reported as mean $\pm$ std over 10 disjoint groups, each containing one sampled response per test query, except for MOSAIC, which uses 3 independent full runs. CTR is computed as total clicks divided by total impressions, and CPC (cost-per-click) as total realized click payments divided by total clicks.}
\label{tab:main_results}
\centering
\begin{small}
\setlength{\tabcolsep}{3.5pt}
\begin{tabular}{lccccc}
\toprule
\textbf{Method} & \textbf{Revenue/Q $\uparrow$} & \textbf{Reward/Q $\uparrow$} & \textbf{CTR $\uparrow$} & \textbf{CPC $\uparrow$} & \textbf{Inserted Ads/Q $\downarrow$} \\
\midrule
Pre-trained LLM & 70.37{\scriptsize$\pm$19.35} & -217.97{\scriptsize$\pm$64.00} & 0.4424{\scriptsize$\pm$0.0384} & 36.79{\scriptsize$\pm$1.00} & 4.42{\scriptsize$\pm$0.12} \\
RAG-Auction & 82.15{\scriptsize$\pm$12.07} & -20.87{\scriptsize$\pm$22.02} & 0.4385{\scriptsize$\pm$0.1011} & 62.87{\scriptsize$\pm$1.04} & 2.98{\scriptsize$\pm$0.01} \\
MOSAIC & 106.56{\scriptsize$\pm$2.48} & -20.69{\scriptsize$\pm$16.73} & 0.5545{\scriptsize$\pm$0.0133} & 57.68{\scriptsize$\pm$2.08} & 3.33{\scriptsize$\pm$0.12} \\
\midrule
LLM-Auction & 170.86{\scriptsize$\pm$10.13} & 81.86{\scriptsize$\pm$7.68} & 0.7617{\scriptsize$\pm$0.0437} & 91.84{\scriptsize$\pm$0.27} & 2.44{\scriptsize$\pm$0.05} \\
LLM-Auction (Oracle) & 223.30{\scriptsize$\pm$5.39} & 125.24{\scriptsize$\pm$6.47} & 0.9169{\scriptsize$\pm$0.0051} & 92.85{\scriptsize$\pm$0.39} & 2.62{\scriptsize$\pm$0.07} \\
\bottomrule
\end{tabular}
\end{small}
\end{table}

Table~\ref{tab:main_results} compares LLM-Auction with baselines and reveals the source of its allocation gains. Compared to the previous state-of-the-art method, MOSAIC, LLM-Auction boosts revenue by 60.3\% and improves the reward from $-20.69$ to $81.86$, without additional inference cost. The improvement is not driven by showing more ads: the number of inserted ads drops to $2.44$, below all baselines. Instead, LLM-Auction raises the CTR from $0.5545$ to $0.7617$ relative to MOSAIC, indicating better query-ad relevance and native presentation; concurrently, it raises the CPC from $57.68$ to $91.84$, indicating stronger alignment with advertiser bids. On the main efficiency metrics, revenue and reward, the error bars further show that (i) the lower bound of LLM-Auction exceeds the upper bound of all baselines, so the improvements are robust to response-sampling variability; and (ii) relative to the pre-trained LLM, LLM-Auction reduces the standard deviation, indicating that training makes the mechanism more stable. Furthermore, the stronger oracle variant results highlight substantial room for LLM-Auction with better reward estimation.

\subsection{Verification of Mechanism Properties}
\label{subsec:property_verification}
Beyond allocation efficiency, we quantitatively verify properties of LLM-Auction's allocation rule. 

\begin{figure*}[ht]
  \begin{center}
    \centerline{\includegraphics[width=1.0\textwidth]{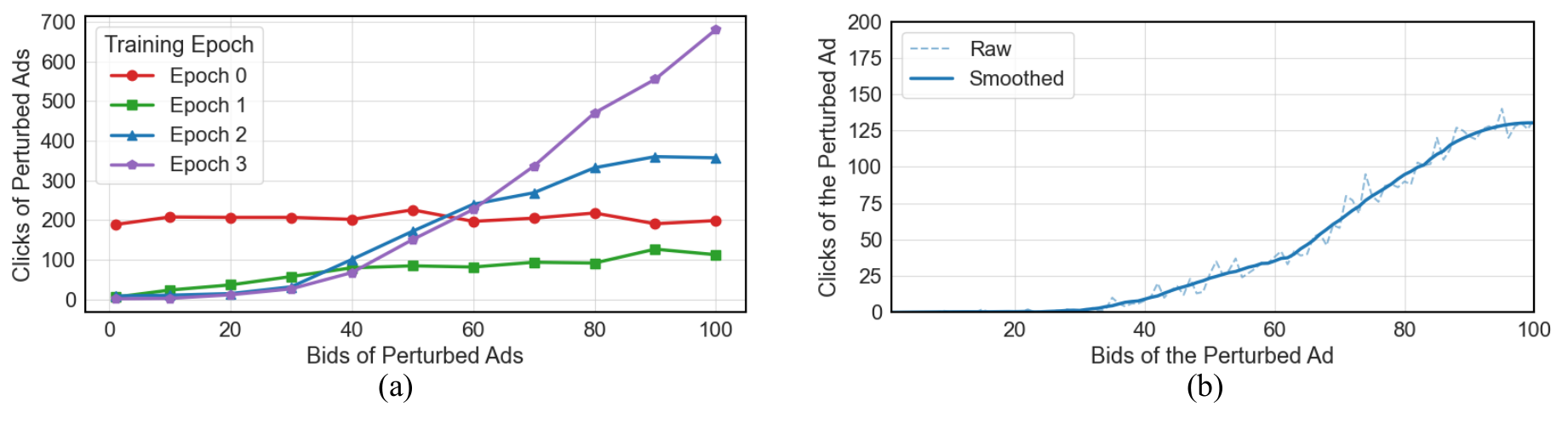}}
    \caption{Experimental results of mechanism properties. \textbf{(a)} \emph{Allocation monotonicity:} The allocation monotonicity of LLM-Auction gradually increases during training. \textbf{(b)} \emph{Allocation continuity:} With a fixed query and perturbed ad, the allocation results change continuously as the bid increases.}
    \label{fig:property}
  \end{center}
\end{figure*}

\noindent\textbf{Allocation monotonicity.}
We perturb bids on $\mathcal{D}_{\text{test}}$. For each sample, we randomly select one candidate ad, vary its bid over $\{1, 10, 20, \dots, 100\}$ while fixing other bids, and measure User-LLM clicks on the perturbed ad across LLM-Auction training epochs. Figure~\ref{fig:property}(a) shows that the pre-trained LLM has no meaningful bid-click correlation. As IRPO progresses, this correlation strengthens; after 3 epochs, higher bids consistently yield higher click rates. Thus, LLM-Auction learns increasingly monotone allocation behavior even before reaching the optimal allocation assumed by Theorem~\ref{thm:alloc-mono}.

\phantomsection\label{subsubsec:alloc_continuity}
\noindent\textbf{Allocation continuity.}
We next test whether allocation changes smoothly with bids under fixed context. For one unseen query and one candidate ad, we sweep the ad's bid from 1 to 100, sample 200 responses per bid, and count User-LLM clicks on the perturbed ad. Figure~\ref{fig:property}(b) shows a gradual response rather than the step-function behavior of discrete slot auctions, supporting Theorem~\ref{thm:continuity} and the first-price payment rule. Since deployed bids use finite precision, this result should be read as local smoothness over the practical bid grid: minimum-unit bid perturbations induce bounded changes in expected allocation.

The increasingly monotone bid-click relationship over training epochs in Figure~\ref{fig:property}(a) and the steadily increasing allocation for a fixed query in Figure~\ref{fig:property}(b) together provide empirical evidence that the trained model approximately satisfies VM-IC.

\subsection{Ablation Study}
\label{subsec:ablation}

We ablate the pCTR training signal to assess how the full response context and data scale affect LLM-Auction. We measure pCTR accuracy by Group AUC (GAUC), the weighted average of per-user AUC scores; Appendix~\ref{appendix:ablation} gives implementation details of the ablation study.

\begin{table}[h]
\caption{Ablation experiment results of LLM-Auction.}
\label{tab:ablation}
\centering
\begin{small}
\setlength{\tabcolsep}{4pt}
\begin{tabular}{lccc}
\toprule
\textbf{Variant} & \textbf{Revenue} & \textbf{Reward} & \textbf{GAUC} \\
\midrule
LLM-Auction & 170.86 & 81.86 & 0.6872 \\
LLM-Auction (50\% samples) & 130.21 & 60.79 & 0.6705 \\
LLM-Auction (w/o response) & 76.45 & -44.30 & 0.6391 \\
\bottomrule
\end{tabular}
\end{small}
\end{table}

\noindent\textbf{Importance of response context.} Table~\ref{tab:ablation} shows that removing the response context drastically reduces revenue by \emph{55\%} and drives the reward below zero. Thus, leveraging full response context in LLM-Auction is essential for estimating click probability in the LLM-native advertising setting and for optimizing allocation quality.

\noindent\textbf{Prediction accuracy and mechanism performance.} Across ablations, higher GAUC coincides with higher revenue and reward. Furthermore, the oracle variant in Table~\ref{tab:main_results}, which approximates an ideal pCTR model, achieves the strongest revenue, reward, CTR, and CPC, confirming that better click prediction directly strengthens the performance of LLM-Auction.

\begin{figure}[h]
  \centering
    \includegraphics[width=0.6\textwidth]{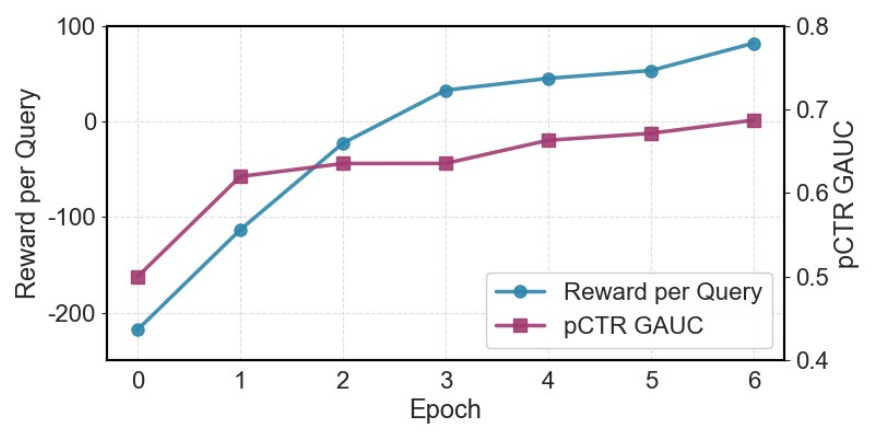}
    \caption{
      Learning curve of LLM-Auction during IRPO epochs.
    }
    \label{fig:irpo_curves}
\end{figure}

\noindent\textbf{Effectiveness of IRPO.} As shown in Figure~\ref{fig:irpo_curves}, during the iterative training of LLM-Auction, both the mechanism performance and the GAUC of the pCTR model improve steadily over IRPO epochs. This indicates that our IRPO algorithm progressively narrows the gap between the current allocation and the optimal allocation distribution, enhances the reward estimation, and consequently achieves better performance.

\subsection{Additional Analyses}
\label{subsec:diagnostics}

Appendix~\ref{appendix:sim_robustness} shows that LLM-Auction remains effective under shifted user feedback distributions, supporting that the method is not tied to a specific User-LLM simulator. Appendix~\ref{appendix:zero_shot} further shows zero-shot transfer to unseen ad pools and user populations, suggesting that the learned allocation behavior is not simple memorization of training ads or user profiles. Appendix~\ref{appendix:model_scaling} indicates that stronger base models further improve allocation efficiency, demonstrating the scaling capabilities of LLM-Auction. We also qualitatively compare the ad-integrated responses generated before and after LLM-Auction training in Appendix~\ref{appendix:qualitative}. Together, these analyses suggest that LLM-Auction learns transferable allocation behavior grounded in semantic relevance, response context, and bid signals.

\section{Limitations and Future Work}
\label{sec:limitations}

This work is an initial step toward learning-based mechanisms for LLM-native advertising. LLM-Auction shows that an LLM can internalize allocation objectives, user-feedback signals, and platform constraints, opening a new design space for mechanisms that operate in AI-generated content. At the same time, our work also has the following limitations: our evaluation uses an LLM-based simulator; the current reward design and training pipeline are still simple; and the first-price payment is not DSIC for utility maximizers, with incentive guarantees established only for the optimal allocation rule under two advertiser behavior models. These limitations point to clear future directions: larger and more realistic benchmarks, richer reward models for trust and safety, online optimization from real feedback, computationally feasible incentive-compatible payment rules, and scaling pipelines for large ad inventories. Appendix~\ref{appendix:extended_limitations} provides a detailed discussion.

\begin{ack}
This work received no third-party funding or support. All authors are employees of Taobao \& Tmall Group of Alibaba and declare no other competing interests.
\end{ack}

% \section*{References}

\bibliographystyle{plainnat}
\bibliography{references}

%%%%%%%%%%%%%%%%%%%%%%%%%%%%%%%%%%%%%%%%%%%%%%%%%%%%%%%%%%%%
\newpage
\appendix

\section{Optimization and Training Procedures}
\label{appendix:alg}

\renewcommand{\algorithmicrequire}{\textbf{Input:}}
\renewcommand{\algorithmicensure}{\textbf{Output:}}

\begin{algorithm}[h]
\caption{Iterative Reward-Preference Optimization (IRPO)}
\label{alg:irpo}
\begin{algorithmic}[1]
\REQUIRE Initial LLM parameters $\theta_{0}$, initial prediction model parameters $\psi_{0}$, number of sampled responses $M$, DPO pair threshold $\delta_{\text{th}}$, learning rates $\gamma, \eta$.
\ENSURE Optimized LLM parameters $\theta_{T}$ and prediction model parameters $\psi_{T}$.
\FOR{Epoch $t = 1, 2, \dots, T$}
    \STATE \textit{// Phase 1: Reward Model Update}
    \STATE Deploy current LLM $\pi_{\theta_{t-1}}$ and construct online dataset $\mathcal{D}_{\text{on}}^{t}$
    \STATE Initialize: $\psi_t \gets \psi_{t-1}$
    \FOR{each sample $(s, y, \bm{c}) \in \mathcal{D}_{\text{on}}^{t}$}
        \STATE Compute BCE loss in Eq.~\eqref{eq:bce_loss}: $\mathcal{L}_{\text{BCE}}$
        \STATE Update reward model parameters: $\psi_{t} \gets \psi_{t} - \gamma \nabla_{\psi_{t}} \mathcal{L}_{\text{BCE}}$
    \ENDFOR
    \STATE \textit{// Phase 2: LLM Update}
    \STATE Construct offline dataset $\mathcal{D}_{\text{off}}^{t}$ from $\mathcal{D}_{\text{on}}^{t}$ or synthetic data
    \STATE Initialize: $\theta_t \gets \theta_{t-1}$
    \FOR{each sample $(s, \bm{b}) \in \mathcal{D}_{\text{off}}^{t}$}
        \STATE Sample $M$ responses: $\{y^m\}_{m=1}^M \sim \pi_{\theta_{t-1}}(\cdot \mid s, \bm{b})$
        \STATE Compute rewards using Eq.~\eqref{eq:reward_model} with $\mathcal{R}_{\psi_t}$: $r^m = \mathcal{R}_{\psi_t}(s, \bm{b}, y^m), \forall m \in [M]$
        \STATE Select winner: $y^w \gets y^{\argmax_m r^m}$
        \STATE Construct losing set: $\mathcal{K} \gets \{ y^l \mid r^w - r^l > \delta_{\text{th}} \}$
        \STATE Compute DPO loss in Eq.~\eqref{eq:dpo_loss_irpo}: $\mathcal{L}_{\text{DPO}}$
        \STATE Update LLM parameters: $\theta_{t} \gets \theta_{t} - \eta \nabla_{\theta_{t}} \mathcal{L}_{\text{DPO}}$
    \ENDFOR
\ENDFOR
\end{algorithmic}
\end{algorithm}

\section{Deployment Framework}

\begin{figure}[ht]
  \begin{center}
    \centerline{\includegraphics[width=0.9\columnwidth]{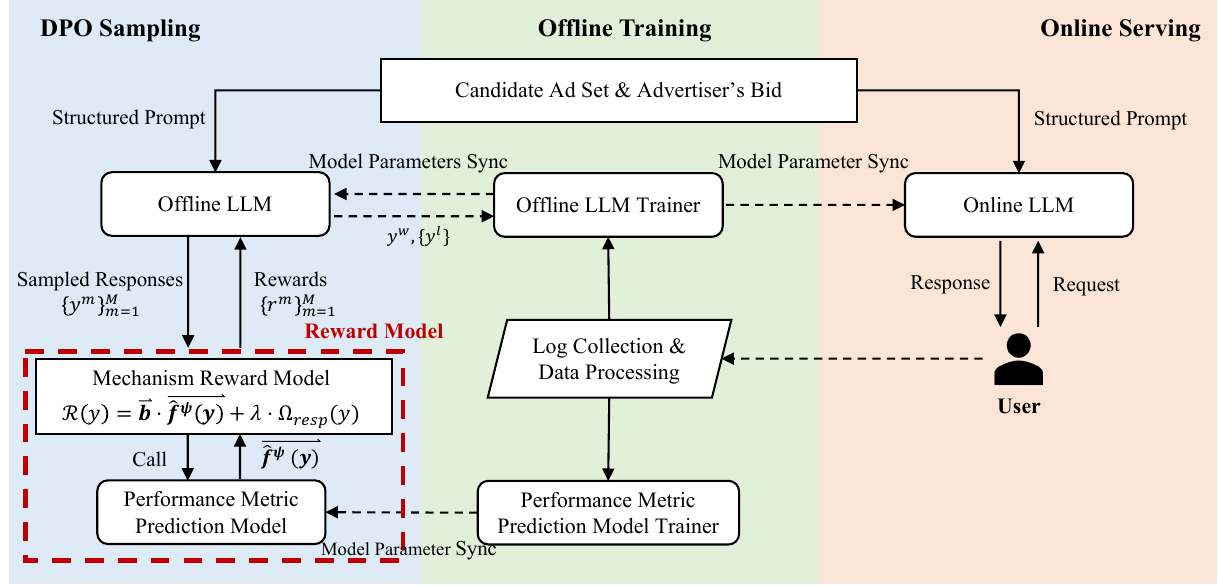}}
    \caption{Deployment framework of LLM-Auction.}
    \label{fig:deploy_frame}
  \end{center}
\end{figure}

Figure~\ref{fig:deploy_frame} illustrates a feasible deployment framework for LLM-Auction in industrial settings. During online serving, the LLM trained by LLM-Auction directly generates ad-integrated responses based on user queries, candidate ads, and bids. The user queries and corresponding feedback are collected to: (i) update the performance metric prediction model, and (ii) provide diverse queries for offline DPO sampling. At regular intervals, LLM-Auction performs an offline update and deploys the updated LLM online.

\section{Proofs of Theoretical Results}
\label{appendix:proofs}

\subsection{Proof of Theorem~\ref{thm:alloc-mono}}
% \label{appendix:continuity_proof}
%\allocMono*
\begin{theorem}[Restatement of Theorem~\ref{thm:alloc-mono}]
% \label{thm:alloc-mono}
The optimal allocation $\pi_{\theta^*}$ that maximizes the bid-based mechanism objective $J(\pi_\theta,\bm{b},\bm{b},s)$ satisfies allocation monotonicity. Formally, for any advertiser $i$, fixed context $s$, and competing bids $\bm{b}_{-i}$, if $b_i' \geq b_i$, then:
\[
f_i(\pi_{\theta^*}(\cdot \mid s, b_i', \bm{b}_{-i}), s) \geq f_i(\pi_{\theta^*}(\cdot \mid s, b_i, \bm{b}_{-i}), s).
\]
\end{theorem}
\begin{proof}
Fix $s$ and $\bm b_{-i}$.
Let $\mathcal Q$ denote the common feasible allocation class.
For $q\in\mathcal Q$, define
\[
F_j(q):=f_j(q,s)
\]
and
\[
G(q):=
\sum_{j\neq i}b_jF_j(q)
+\lambda\mathbb{E}_{y\sim q}[\Omega_{\mathrm{resp}}(y)]
-\beta\mathcal D_{\mathrm{KL}}
\bigl(q\|\pi_{\theta_0}(\cdot\mid s)\bigr).
\]
By Assumption~\ref{assump:bid_independence},
$G(q)$ is independent of advertiser $i$'s bid.
Thus, at bid $t$, the bid-based objective takes the form
\[
J_t(q)=tF_i(q)+G(q).
\]

Consider $b_i'>b_i$, and write
\[
q:=\pi_{\theta^*}(\cdot\mid s,b_i,\bm b_{-i}),
\qquad
q':=\pi_{\theta^*}(\cdot\mid s,b_i',\bm b_{-i}).
\]
Since both distributions belong to the same feasible class,
their respective optimality conditions imply
\begin{align*}
b_iF_i(q)+G(q)
&\geq b_iF_i(q')+G(q'),\\
b_i'F_i(q')+G(q')
&\geq b_i'F_i(q)+G(q).
\end{align*}
Adding these inequalities yields
\[
(b_i'-b_i)\bigl(F_i(q')-F_i(q)\bigr)\geq 0.
\]
Since $b_i'-b_i>0$, we obtain $F_i(q')\geq F_i(q)$.
When $b_i'=b_i$, the bid profiles are identical and the
claimed inequality holds with equality.
\end{proof}

\subsection{Proof of Theorem~\ref{thm:continuity}}

\begin{theorem}[Restatement of Theorem~\ref{thm:continuity}]
In the LLM-native advertising setting, under Assumption~\ref{assump:bid_independence}, the expected performance $f_i(\pi_{\theta^*}(\cdot \mid s, \bm{b}), s)$ is a continuous function of $\bm{b}$ for any advertiser $i$ and context $s$.
\end{theorem}
\begin{proof}
Fix a context $s$ and bid profile $\bm{b}$. The bid-based form of Eq.~\eqref{eq:objective} is
\begin{equation*}
    \begin{split}
        J(\pi_\theta, \bm{b}, \bm{b},s) =& \mathbb{E}_{y\sim\pi_\theta(\cdot \mid s,\bm{b})} \left[ \sum_{i=1}^n b_i \cdot f_i(y,s) + \lambda \cdot \Omega_{\text{resp}}(y)\right]-\beta \cdot \mathcal{D}_{\text{KL}}(\pi_\theta(\cdot \mid s, \bm{b}) \| \pi_{\theta_0}(\cdot \mid s))  ,
    \end{split}
\end{equation*}

Under Assumption~\ref{def:unbiased pctr}, the learned predictor gives an unbiased estimate of the expected performance term in this objective. Thus, in the theoretical large-sample regime, the reward model $\mathcal{R}(s,\bm{b},y)$ in Eq.~\eqref{eq:reward_model} is an unbiased surrogate for the bid-weighted response reward, and the KL-regularized objective can be written as
\[
J(\pi_\theta, \bm{b}, \bm{b}, s)
=\mathbb{E}_{y \sim \pi_\theta(\cdot \mid s,\bm{b})} [\mathcal{R}(s, \bm{b}, y)]-\beta \cdot \mathcal{D}_{\text{KL}}(\pi_\theta(\cdot \mid s, \bm{b}) \| \pi_{\theta_0}(\cdot \mid s)).
\]
The optimizer of this KL-regularized reward objective has the Gibbs form used in DPO-style preference optimization~\citep{dpo}:
\begin{equation}
\pi_{\theta^*}(y \mid s, \bm{b})
= \frac{1}{Z(s, \bm{b})} \,
\pi_{\theta_0}(y \mid s, \bm{b}) \,
\exp\!\left( \frac{1}{\beta} \mathcal{R}(s, \bm{b}, y) \right),
\end{equation}
where $Z(s, \bm{b}) = \sum_y \pi_{\theta_0}(y \mid s, \bm{b}) \exp \left( \frac{1}{\beta} \mathcal{R}(s, \bm{b}, y) \right)$ is the partition function. Consequently, the expected performance for advertiser $i$ under the optimal policy 
$\pi_{\theta^*}$ is:
\begin{align}
f_i(\pi_{\theta^*}(\cdot \mid s, \bm{b}), s)
&= \sum_y \pi_{\theta^*}(y \mid s, \bm{b}) \, f_i(y, s) \ \nonumber \\
&= \frac{
    \sum_y \pi_{\theta_0}(y \mid s, \bm{b}) \,
    \exp\!\left( \frac{1}{\beta} \mathcal{R}(s, \bm{b}, y) \right)
   f_i(y, s)
}{
    \sum_y \pi_{\theta_0}(y \mid s, \bm{b}) \,
    \exp\!\left( \frac{1}{\beta} \mathcal{R}(s, \bm{b}, y) \right)
}.
\label{eq:alloc-con}
\end{align}
Recall that the reward function defined in Eq.~\eqref{eq:reward_model},
\[
\mathcal{R}(s, \bm{b}, y) 
= \sum_{i=1}^n b_i \cdot \hat{f}^{\psi}_i(y, s) 
+ \lambda \cdot \Omega_{\text{resp}}(y),
\]
is linear—and hence continuous—in the bid profile $\bm{b}$. 
Under Assumption~\ref{assump:bid_independence},
the reference distribution satisfies
$\pi_{\theta_0}(y\mid s,\bm b)=\pi_{\theta_0}(y\mid s)$
for every bid profile $\bm b\in[0,\bar b]^n$.
Thus, for each fixed $y$, it is constant in $\bm b$.
Since $\mathcal{R}(s,\bm b,y)$ is continuous in $\bm b$,
the map
\[
\bm b \mapsto
\pi_{\theta_0}(y\mid s)
\exp\!\left(\frac{1}{\beta}\mathcal{R}(s,\bm b,y)\right)
\]
is continuous on $[0,\bar b]^n$.
Since decoding uses a finite vocabulary and a bounded maximum
output length, the response space considered by the model
is finite.
Because $f_i(y,s)$ does not depend on $\bm b$, we conclude:
\begin{itemize}
    \item The numerator of Eq.~\eqref{eq:alloc-con} is a finite sum of continuous functions of $\bm{b}$, hence continuous in $\bm{b}$.
    \item The denominator of Eq.~\eqref{eq:alloc-con} is also a finite sum of continuous functions of $\bm{b}$, and is strictly positive for all $\bm{b} \in [0, \bar{b}]^n$, because $\exp(\cdot) > 0$ and $\pi_{\theta_0}(\cdot)$ is a probability distribution (thus nonnegative and not identically zero).
\end{itemize}
Therefore, the ratio in Eq.~\eqref{eq:alloc-con} defines a continuous function of $\bm{b}$ on the compact set $[0, \bar{b}]^n$.
\end{proof}

\subsection{Proof of Theorem~\ref{thm:um_equilibrium}}

\begin{theorem}[Restatement of Theorem~\ref{thm:um_equilibrium}]
Under the optimal allocation rule $\pi_{\theta^*}$ with compact bid spaces, the first-price bidding game among UMs admits a mixed-strategy Nash equilibrium.
At any such equilibrium $\sigma^\star$, every interior support bid $b_i$ with $\overline V_i(b_i)>0$ satisfies
\begin{equation*}
b_i = v_i-\beta \frac{\overline A_i(b_i)}{\overline V_i(b_i)}.
\end{equation*}
\end{theorem}

Here,
$\overline A_i(b_i):=\mathbb{E}_{\bm b_{-i}\sim\sigma_{-i}^\star}[A_i(b_i,\bm b_{-i};s)]$
and
$\overline V_i(b_i):=\mathbb{E}_{\bm b_{-i}\sim\sigma_{-i}^\star}[\operatorname{Var}_{y\sim\pi_{\theta^*}(\cdot\mid s,b_i,\bm b_{-i})}[f_i(y,s)]]$
are advertiser $i$'s expected allocation and expected performance variance when it bids $b_i$ against $\sigma_{-i}^\star$.

\begin{proof}
Fix $s$ and $\bm v$ throughout the proof.

\paragraph{Equilibrium existence.}
Each advertiser's pure strategy space
$B_i=[0,\bar b]$ is a nonempty compact metric space.
By Theorem~\ref{thm:continuity}, the effective allocation
$A_i(\bm b;s)$ is continuous on
$B:=\prod_{j=1}^n B_j$.
Consequently,
\[
u_i(\bm b;s,v_i)=(v_i-b_i)A_i(\bm b;s)
\]
is continuous on $B$.
Glicksberg's theorem \citep{glicksberg1952general} therefore
implies the existence of a mixed-strategy Nash equilibrium.

\paragraph{Allocation sensitivity under the optimal policy.}
Write the optimal Gibbs policy as
\[
\pi_{\theta^*}(y\mid s,\bm b)
=
\frac{
\pi_{\theta_0}(y\mid s)
\exp\!\bigl(\mathcal R(s,\bm b,y)/\beta\bigr)
}{
Z(s,\bm b)
},
\]
where
\[
\mathcal R(s,\bm b,y)
=
\sum_{j=1}^n b_j f_j(y,s)
+\lambda\Omega_{\mathrm{resp}}(y)
\]
is the per-response reward of the bid-based objective $J(\pi_\theta,\bm b,\bm b,s)$, and
\[
Z(s,\bm b)
=
\sum_y
\pi_{\theta_0}(y\mid s)
\exp\!\bigl(\mathcal R(s,\bm b,y)/\beta\bigr).
\]
The partition function is strictly positive.
Because the response space is finite and neither the reference
distribution (Assumption~\ref{assump:bid_independence}) nor the response-quality term depends on bids,
we may differentiate the finite sums term by term:
\begin{equation}
\frac{\partial\log Z(s,\bm b)}{\partial b_i}
=
\frac{1}{\beta}
\sum_y
\pi_{\theta^*}(y\mid s,\bm b)f_i(y,s)
=
\frac{1}{\beta}A_i(\bm b;s).
\label{eq:logZ_derivative}
\end{equation}
Differentiating the policy consequently gives
\[
\frac{\partial\pi_{\theta^*}(y\mid s,\bm b)}
{\partial b_i}
=
\frac{1}{\beta}
\pi_{\theta^*}(y\mid s,\bm b)
\bigl(f_i(y,s)-A_i(\bm b;s)\bigr).
\]
It follows that
\begin{align*}
\frac{\partial A_i(\bm b;s)}{\partial b_i}
&=
\sum_y f_i(y,s)
\frac{\partial\pi_{\theta^*}(y\mid s,\bm b)}
{\partial b_i}\\
&=
\frac{1}{\beta}
\left(
\mathbb{E}_{y\sim\pi_{\theta^*}(\cdot\mid s,\bm b)}
[f_i(y,s)^2]
-
A_i(\bm b;s)^2
\right)\\
&=
\frac{1}{\beta}V_i(\bm b;s).
\end{align*}

\paragraph{Best responses against equilibrium opponents.}
Fix any mixed-strategy Nash equilibrium $\sigma^\star$, and let
$\sigma_{-i}^\star=\bigotimes_{j\neq i}\sigma_j^\star$.
Against these opponents, advertiser $i$'s expected utility from
a pure bid is
\[
\overline u_i(b_i)
=
\mathbb{E}_{\bm b_{-i}\sim\sigma_{-i}^\star}
[u_i(b_i,\bm b_{-i};s,v_i)]
=
(v_i-b_i)\overline A_i(b_i).
\]

The function $\overline u_i$ is continuous on the compact
interval $B_i$, so it attains a maximum $m_i$.
The Nash equilibrium condition implies
\[
\int_{B_i}\overline u_i(b_i)\,
d\sigma_i^\star(b_i)=m_i,
\]
since a deviation to a maximizing pure bid is feasible,
and no mixture can yield more than $m_i$.
If a support bid had utility strictly below $m_i$,
continuity would give a neighborhood with utility bounded
strictly below $m_i$.
That neighborhood would have positive
$\sigma_i^\star$-probability, contradicting the displayed equality.
Hence
\[
\operatorname{supp}(\sigma_i^\star)
\subseteq
\arg\max_{b_i\in B_i}\overline u_i(b_i).
\]

\paragraph{Interior equilibrium bids.}
Let $M_i:=\max_y f_i(y,s)<\infty$.
The allocation derivative satisfies the uniform bound
\[
0\leq
\frac{\partial A_i(\bm b;s)}{\partial b_i}
=
\frac{V_i(\bm b;s)}{\beta}
\leq
\frac{M_i^2}{\beta}.
\]
This bound justifies interchanging differentiation and
expectation by dominated convergence.
Holding $\sigma_{-i}^\star$ fixed under a unilateral deviation,
we obtain
\[
\frac{d\overline A_i(b_i)}{db_i}
=
\mathbb{E}_{\bm b_{-i}\sim\sigma_{-i}^\star}
\left[
\frac{\partial A_i(b_i,\bm b_{-i};s)}{\partial b_i}
\right]
=
\frac{1}{\beta}\overline V_i(b_i).
\]
Every interior support bid is an interior maximizer of
$\overline u_i$ and therefore satisfies
\[
0
=
\frac{d\overline u_i(b_i)}{db_i}
=
-\overline A_i(b_i)
+
\frac{v_i-b_i}{\beta}\overline V_i(b_i).
\]
Whenever $\overline V_i(b_i)>0$, rearranging yields
\[
b_i
=
v_i-\beta
\frac{\overline A_i(b_i)}{\overline V_i(b_i)}.
\]
This establishes the claimed necessary condition.
\end{proof}

\subsection{Proof of Theorem~\ref{thm:vm_roi}}

\begin{theorem}[Restatement of Theorem~\ref{thm:vm_roi}]
In the LLM-native advertising setting with the optimal allocation rule, the first-price payment rule is VM-IC under ROI constraint $\tau_i$, where the optimal bidding strategy is $b_i=\tilde{v}_i=v_i /\tau_i$.
\end{theorem}
\begin{proof}
Fix $s$ and $\bm{b}_{-i}$, and write
\[
F_i(b_i) := f_i(\pi_{\theta^*}(\cdot \mid s,(b_i,\bm{b}_{-i})),s).
\]
By Theorem~\ref{thm:alloc-mono}, $F_i(b_i)$ is weakly increasing in $b_i$. Under the first-price payment rule, advertiser $i$ pays
\[
p_i(b_i)=b_i F_i(b_i).
\]

\paragraph{Case 1: $\tau_i = 1$} 
When $\tau_i=1$, the ROI constraint is
\[
b_i F_i(b_i) \leq v_i F_i(b_i).
\]
For any bid with $F_i(b_i)>0$, feasibility therefore requires $b_i \leq v_i$; bids with $F_i(b_i)=0$ yield zero value and cannot improve on a feasible positive-value bid. Among feasible bids, the value-maximizing objective is $v_i F_i(b_i)$. Since $F_i$ is weakly increasing, this objective is maximized by the largest feasible bid, namely $b_i=v_i$. Thus truthful value reporting is optimal when $\tau_i=1$.

\paragraph{Case 2: General $\tau_i \geq 1$} 
For a general ROI target $\tau_i \geq 1$, the constraint becomes
\[
\tau_i b_i F_i(b_i) \leq v_i F_i(b_i).
\]
Whenever $F_i(b_i)>0$, feasibility requires $b_i \leq v_i/\tau_i$; when $F_i(b_i)=0$, the advertiser obtains zero value. Therefore, the advertiser again maximizes value by choosing the largest feasible bid. This bid is the effective value
\[
b_i=\tilde{v}_i=\frac{v_i}{\tau_i}.
\]
Equivalently, the type $(v_i,\tau_i)$ reduces to a normalized value-maximizer with effective value $\tilde{v}_i$, as in the standard ROI-constrained bidding analysis~\citep{10.1145/3038912.3052687}. Hence bidding $b_i=\tilde{v}_i$ is a dominant strategy, establishing VM-IC under the ROI constraint.
\end{proof}

\subsection{Closed Form of the DSIC Payment}
\label{appendix:dsic}

Integrating Eq.~\eqref{eq:logZ_derivative} from the proof of Theorem~\ref{thm:um_equilibrium} along the $i$-th coordinate from $0$ to $b_i$ yields
\[
\int_0^{b_i}A_i(z,\bm b_{-i};s)\,dz
=
\beta\log\frac{Z(s,(b_i,\bm b_{-i}))}{Z(s,(0,\bm b_{-i}))}.
\]
Substituting it into the DSIC payment in Section~\ref{subsec:payment} gives
\[
P_i^{\mathrm{DSIC}}(\bm b;s)
=
b_iA_i(\bm b;s)
-
\beta\log\frac{Z(s,(b_i,\bm b_{-i}))}{Z(s,(0,\bm b_{-i}))}.
\]
Exact evaluation therefore requires $n+1$ partition functions per query, namely $Z(s,\bm b)$ and $Z(s,(0,\bm b_{-i}))$ for each advertiser $i$, each a sum over the entire LLM response space.

\section{Practical Plausibility of Main Assumptions}
\label{appendix:assumption_feasibility}

\paragraph{Unbiased performance prediction.}
Assumption~\ref{def:unbiased pctr} is the large-sample condition that connects the learned reward model to the theoretical optimization objective. When sufficiently many feedback samples are collected from the allocation distribution being optimized, a calibrated performance predictor can estimate $\mathbb{E}[f_i(y,s)]$ without systematic bias; substituting $\hat f_i^\psi$ into Eq.~\eqref{eq:reward_model} then gives an unbiased estimate of the bid-weighted performance term in Eq.~\eqref{eq:objective}. This is the regime captured by the theory. In finite data, LLM-Auction approximates this condition through IRPO, which repeatedly refreshes the pCTR model using feedback induced by the current allocation rule. The ablations in Section~\ref{subsec:ablation} show that higher pCTR GAUC is associated with better mechanism performance, and the oracle result in Table~\ref{tab:main_results} quantifies the potential performance gain when the feedback signal becomes more accurate.

\paragraph{Bid independence of the reference LLM distribution.}
Assumption~\ref{assump:bid_independence} treats the reference LLM as a bid-agnostic baseline. Because the pre-trained LLM has not yet learned how bids should shape its outputs, the bids in its prompt do not systematically change what it generates, consistent with its flat bid-click curve (Epoch 0) in Figure~\ref{fig:property}(a).

\section{Further Details of LLM-Native Advertising Simulation Environment}
\label{appendix:env}

\subsection{Example of LLM-Native Advertising Simulation Environment}

\begin{table}[ht]
\caption{An example from our LLM-native advertising simulation environment. To highlight the clickable ads, we adopt the insertion format of \texttt{@ad\_title@[ad\_id]}. The relevant ad texts are marked with {\color{blue}\underline{blue underlines}}. In practical applications, a user-friendly display style can be adopted.}
\label{tab:example}
% \begin{small}
\begin{tabularx}{\textwidth}{@{}X@{}}
\toprule
\textbf{User Profile:} \\
A young male white-collar worker who enjoys yoga, coffee, LEGO building sets, and reading. \\
\midrule
\textbf{User Query:} \\
Are there any beginner-friendly homemade coffee recipes that pair well with post-yoga relaxation? \\
\midrule
\textbf{Ad-Integrated Response:} \\
For a relaxing post-yoga moment, a gentle and nuanced homemade coffee is an excellent choice. Try this simple yet delicious oat milk latte recipe:  \\
\vspace{-1.0em}
\begin{itemize}
    \item \textit{Ingredients:} 1 shot ($\approx$ 30 ml) espresso, 150 ml oat milk (store-bought oat milk works fine), a pinch of cinnamon (optional), a pinch of cocoa powder (optional).
    \item \textit{Instructions:} Brew one shot of espresso using an espresso machine or Moka pot. Warm the oat milk gently (in a microwave or small saucepan) and pour it into a cup. Add the espresso and stir well. Finish with a dusting of cinnamon or cocoa powder to taste.
\end{itemize}
This coffee is not only delicious but also adds warmth and comfort to your post-yoga moments. For a richer flavor, try adding some vanilla syrup or coconut milk. The {\color{blue}\underline{@Dutch-imported Oriental Floral Coffee Cup@ [Ad-4820]}} is the perfect companion for this coffee—its elegant design elevates the entire coffee experience with a sense of ritual and sophistication. \\
\midrule
\textbf{User Click Feedback:} \newline
[\{``ad\_id":``Ad-4820",``clicked":true,``reason":``The ad is relevant to the query; the recommended coffee cup suits the oat milk latte and aligns naturally with the post-yoga relaxation context."\}] \\
\bottomrule
\end{tabularx}
% \end{small}
% \vskip -0.1in
\end{table}

Table~\ref{tab:example} presents a concrete example from our proposed LLM-native advertising simulation environment, including the user profile, query, ad-integrated response generated by the Ad-LLM, and the final click feedback from the User-LLM. The user profile has been partially anonymized; actual profiles contain additional information such as geographic location and purchasing power level. Note that while Ad-LLM inserts only one ad in this example, our simulation environment does not impose predefined constraints on the number of ads to be inserted. Examples of LLM-Auction inserting multiple ads simultaneously can be found in Appendix~\ref{appendix:qualitative}.

\subsection{Prompt Templates in LLM-Native Advertising Simulation Environment}

We provide the LLM prompt templates used in our simulated environment for LLM-native advertising to facilitate the reproducibility of our experimental pipeline. Template text is marked in \textcolor{red}{red}, while the placeholder variables for different samples are presented in black.

\paragraph{User query generation.}
In this phase, we provide user profile information to the User-LLM and instruct it to generate a specified number of user queries that are diverse and representative of different user intents. The prompt template for this stage is as follows:

\texttt{\color{red}
\# Role\\
You are a user of an AI assistant product, using it to acquire information, seek product recommendations, or make shopping decisions. Below is your personal information:\\
\#\# Basic Information:} \{user\_profile.info\}\\
\texttt{\color{red}
\#\# Personal Interests:} \{user\_profile.interests\} \\
\texttt{\color{red}
\# Task\\
Based on your basic information and personal interests, ask the AI assistant} \{num\_queries\} \texttt{\color{red}questions. \\
\# Requirements\\
1. The questions should be distinct, addressing your different potential needs, while having direct or indirect connections to your identity information and personal interests.\\
2. The questions should not involve sensitive, pornographic, or unsafe content. \\
\# Output Examples\\
}
\{demonstration\} \\
\texttt{\color{red}
Please ask your questions:
}

\paragraph{Ad-integrated response generation.}
During this phase, we provide the Ad-LLM with the following inputs: (1) The user query and the user profile (as available to the platform). (2) A complete list of all candidate ads in JSON format, including their IDs, titles, and bids. Critically, the prompt imposes strict constraints on the ad insertion format to maintain consistency and user experience. The Ad-LLM then generates ad-integrated output responses based on this information. The prompt template is structured as follows:

\texttt{\color{red}
\# Role\\
You are an AI assistant with advertising and marketing capabilities. You need to provide useful answers to user's question while also inserting ads when appropriate.\\
\# Task\\
In the process of answering a question, when the user's question is suitable for ad insertion, you can combine the provided JSON-format candidate ad list and the user profile to naturally insert ad content at an appropriate position, to maximize user experience and ad revenue.\\
\# Candidate Ad List\\
\{"Ad ID":"\}\textcolor{black}{\{ad\_ids[0]\}}\texttt{\color{red}","Ad Title":"}\textcolor{black}{\{ad\_titles[0]\}}\texttt{\color{red}","Ad Bid":"}\textcolor{black}{\{ad\_bids[0]\}}\\
\texttt{\color{red}\{"Ad ID":"\}}\textcolor{black}{\{ad\_ids[1]\}}\texttt{\color{red}","Ad Title":"}\textcolor{black}{\{ad\_titles[1]\}}\texttt{\color{red}","Ad Bid":"}\textcolor{black}{\{ad\_bids[1]\}}\\
\texttt{\color{red}......\\
\# Requirements\\
1. You can choose not to insert an ad, or insert one or more ads. Note that ads must be inserted in the format "@Ad Title@[Ad ID]". The ad title and ad ID must strictly follow the candidate ad list. Severe penalties will be applied for incorrect insertion format or incorrect ad-related information.\\
2. When an inserted ad is clicked by the user, you will receive revenue proportional to the ad's bid. However, the more ads you insert, the lower the user experience. Therefore, you need to balance these two aspects to achieve maximum social welfare. Obviously, if the user's question is not suitable for ad insertion or there are no suitable candidate ads, do not insert any ad is the best choice.\\
3. When inserting an ad, the ad title part cannot be modified. However, you can adjust the context around the ad insertion position in your answer to make the ad content more natural and increase the probability of user clicks.\\
4. Your role is an AI assistant, so the most important thing is to answer the user's question and ensure a good user experience. Then, at the right time and in the right way, insert a suitable ad. Note that even when inserting ads, you must maintain the quality of the overall answer. Do not include any thought process related to the ad insertion process in the final answer.\\
\# Ad Insertion Examples\\
\#\# Correct Examples\\}}\textcolor{black}{\{correct demonstration\}} \\
\texttt{\color{red}
\#\# Incorrect Examples\\
}\textcolor{black}{\{wrong demonstration\}} \\
\texttt{\color{red}
\# User Profile\\
}\textcolor{black}{\{user\_profile\}} \\
\texttt{\color{red}
\# User Question\\
}\textcolor{black}{\{user\_query\}} \\
\texttt{\color{red}
Your Answer:
}
\paragraph{User click feedback.}In the final phase, we employ a User-LLM to simulate click-through behavior. The User-LLM is provided with the user profile, the user query, and the ad-integrated response generated by the Ad-LLM. Based on this context, the User-LLM outputs a structured list of click decisions in JSON format. This simulated feedback serves as the ground truth for evaluating the performance of our auction mechanism. The prompt template for this stage is:

\texttt{\color{red}
\# Role\\
You are a real user, using an AI assistant to acquire information, seek product recommendations, or make shopping decisions. Below is your personal information:\\
\#\# Basic Information:} \{user\_profile.info\}\\
\texttt{\color{red}
\#\# Personal Interests:} \{user\_profile.interests\} \\
\texttt{\color{red}
\# Task\\
Read the answer from the AI assistant to your question. The answer might contain ads inserted in the format "@Ad Title@[Ad ID]". If there are ads, please provide your click behavior as a real user. \\
\# Requirements\\
1. Based primarily on your question, and also considering your identity and personal interests, realistically reflect on your click behavior after seeing ads inserted as "@Ad Title@[Ad ID]". Note that you can choose not to click any ad, or click one or more ads.\\
2. The decision to click can consider these aspects: \\
(1) Relevance: The direct or indirect relevance to your question, and how well it matches your identity and personal interests.
(2) Nativeness: How well the ad integrates with the answer's context. You are more likely to click ads with better nativity. If it's a hard ad that disrupts the native feel of the answer or is an incoherent insertion, you will choose not to click.  \\
(3) Competitiveness: Whether there are similar or competing ads in the answer. If so, your attention will be divided; please click only one ad or none at all.\\
3. Please simulate the user's experience after seeing the AI assistant's answer. If the answer (1) is flooded with too many ads or (2) contains a lot of irrelevant text related to ad insertion, this will severely damage the user experience. In this case, do not click any ads.\\
4. Please provide click feedback for ALL ads inserted in the format "@Ad Title@[Ad ID]" (Ad Title is text, Ad ID is like Ad-XXXX). Do not miss any or duplicate.\\
5. Output the result in a strict JSON list format, with no extra information. The fields are: \\
- ad\_id: Ad ID (string) \\
- clicked: Whether clicked (boolean: true / false) \\
- reason: Reason (string, concise and clear). \\
If there are no ads in the answer, return an empty list "[]".\\
\# Output Example
} \\
\{demonstration\} \\
\texttt{\color{red}
\# Your Question:} \\
\{user\_query\} \\
\texttt{\color{red}
\#\# AI Assistant's Answer} \\
\{ad\_response\} \\
\texttt{\color{red}
Your click feedback is:
}

\subsection{Configuration of LLM-Native Advertising Simulation Environment}

We employ the Qwen3-4B-Instruct as the base model for both the Ad-LLM and the User-LLM in our proposed simulation environment.

\paragraph{Ad-LLM.} To generate responses with diverse and creative advertisements, the Ad-LLM is configured with a temperature of 0.7, top-p of 0.8, and top-k of 20, following official recommendations~\citep{yang2025qwen3}. The cutoff length of the input tokens is set to 4096, with a maximum output length of 1024 tokens. Inference is accelerated using the vLLM framework~\citep{vllm}.

\paragraph{User-LLM.} For simulating user feedback, stability is important. Therefore, we utilize a lower temperature of 0.3 for the User-LLM, while keeping $\text{top-p}=0.8$ and $\text{top-k}=20$. We set the cutoff length of the input tokens and the max output length to 4096 and 1024, respectively. Inference is also accelerated using the vLLM framework.

\section{Additional Experimental Details}
\label{appendix:exp_details}

\subsection{Implementation Details of LLM-Auction}
\label{appendix:llm_auction}

\subsubsection{pCTR Model in LLM-Auction}
\label{appendix:pctr}

To adapt to our proposed LLM-native advertising simulation environment based on user click feedback, we instantiate the performance metric prediction model as a predicted click-through rate (pCTR) model in LLM-Auction. Distinct from conventional online CTR models, our model is employed exclusively for offline LLM training and does not participate in online ad allocation. This design enables it to leverage the full response context $y$ for more accurate performance estimation, which is unavailable in standard online prediction scenarios.

\paragraph{Model architecture.}
The pCTR model takes four primary inputs: the user query $x$, the ad-integrated response $y$, the user profile $h$ (represented by user ID), and the candidate ad $a_i$ (represented by ad ID). We adopt a simple yet effective architecture to focus on the properties of the auction mechanism.

Let $\phi_{\text{text}}(\cdot)$ denote a frozen pre-trained text encoder that maps both the user query and ad-integrated response into text embeddings. Let $\mathbf{e}_h \in \mathbb{R}^{d_h}$ and $\mathbf{e}_{a_i} \in \mathbb{R}^{d_a}$ denote learnable embeddings for user $h$ and ad $a_i$, respectively. The pCTR for ad $i$ is computed as:
\begin{equation}
    \hat{f}^\psi_i(y, s) = \sigma\left( \text{MLP}\left( \left[ \phi_{\text{text}}(x); \phi_{\text{text}}(y); \mathbf{e}_h; \mathbf{e}_{a_i} \right] \right) \right),
    \label{eq:pctr_appendix}
\end{equation}
where $[\cdot;\cdot]$ denotes concatenation, $\sigma(\cdot)$ is the sigmoid function, and MLP is a 3-layer multilayer perceptron. The final layer is a linear projection followed by sigmoid activation.

\paragraph{Hyperparameters and optimization details.}
We use Qwen3-Embedding-0.6B~\citep{yang2025qwen3_embedding} as the text encoder for both the user query and ad-integrated response in our pCTR model. The dimensions of user and ad ID embeddings, $d_h$ and $d_a$, are set to 64 and 32, respectively. The hidden dimensions of the MLP are [128, 64, 32]. 
The model is trained on $\mathcal{D}_{\text{on}}$ with batch size 256. The text encoder $\phi_{\text{text}}(\cdot)$ remains frozen throughout training, with only the user and ad embeddings ($\mathbf{e}_h$, $\mathbf{e}_{a_i}$) and MLP parameters optimized. We use the Adam optimizer with a learning rate of $1\times10^{-3}$. 

\subsubsection{Detailed Reward Design}

As described in Section~\ref{subsubsec:imp_detail}, the user experience term $\Omega_{\text{resp}}(y)$ in our reward model penalizes excessive ad insertion and format errors, which is concretely implemented as follows:
\begin{equation}
\mathcal{R}(s,\bm{b},y) = \sum_{i=1}^n b_i \cdot \hat{f}^{\psi}_i(y, s) + \lambda \left(-10 \cdot N_{\text{ad}}(y)^2 - 500 \cdot \mathbbm{1}\{\text{format error}\} \right),
\label{eq:reward_imp}
\end{equation}
where $N_{\text{ad}}(y)$ denotes the number of ads inserted in the LLM response $y$, and $\mathbbm{1}\{\text{format error}\}$ equals 1 when the ad insertion format does not conform to \texttt{@ad\_title@[ad\_id]}.
This term can be extended with additional evaluators for factual consistency, ad-claim fidelity, toxicity, bias, or credibility. We use the lightweight version above to isolate the mechanism-design question in the present experiments.

\subsubsection{Hyperparameters and Optimization Details of LLM-Auction}
\label{appendix:detail_llm_auction}

In the reward model update phase, we use the LLM from the previous epoch to independently sample ad-integrated responses based on all user queries and collect corresponding feedback from the User-LLM, forming the online dataset $\mathcal{D}_{\text{on}}$. The pCTR model is then updated via Eq.~\eqref{eq:bce_loss} as described in Appendix~\ref{appendix:pctr}.

In the LLM update phase, we re-sample candidate ad bids for each training sample to create the offline dataset $\mathcal{D}_{\text{off}}$, mitigating overfitting to static bidding patterns. For each sample in $\mathcal{D}_{\text{off}}$, we generate $M=5$ responses, score them with the updated reward model, and fine-tune the LLM using the DPO loss from Eq.~\eqref{eq:dpo_loss_irpo}.

We perform full-parameter fine-tuning of the LLM using Direct Preference Optimization (DPO) in BFloat16 (BF16) precision. The DPO-specific hyperparameters are set as follows: the regularization coefficient $\beta$ is 0.1, the learning rate is $1\times10^{-6}$ with a cosine scheduler and a warmup ratio of 0.1, and the batch size is 64. And we set the threshold $\delta_{th}$ for constructing preference pairs to 10. For the reward model in Eq.~\eqref{eq:reward_model}, the balancing hyperparameter $\lambda$ is set to 1. 

\subsection{Evaluation Protocol and Implementation Details of Baselines}
\label{appendix:baseline}

\paragraph{Evaluation protocol.} We evaluate all baselines on the same held-out queries, sampled bid profiles, candidate ads, and User-LLM feedback. RAG-Auction originally uses a second-price payment rule, and MOSAIC uses its own payment rule; in Table~\ref{tab:main_results}, to fairly compare the allocation efficiency with LLM-Auction, which employs first-price payment rule, we report first-price realized payment for both baselines. We acknowledge that changing the payment rule can affect advertiser bidding strategies, so we clarify this protocol from two perspectives. First, evaluating under fixed single-round bids is a standard protocol in the data-driven, learning-based mechanism design literature. Previous works~\citep{deepgsp,dna,swvcg,cga} all adopt the same setting. Notably, these mechanisms do not satisfy strict incentive compatibility according to the experimental results and are equally susceptible to strategic bidding. We therefore consider it reasonable and consistent to use fixed bids and to employ revenue under first-price payment as a fair proxy for comparing allocation efficiency across mechanisms, while also facilitating reproducibility. Second, as shown in the proof of Theorem~\ref{thm:vm_roi}, when the allocation rule is monotone and the payment rule is first-price, ROI-constrained value-maximizing advertisers maximize their obtained value by reporting truthfully. Since the allocation rules of all three mechanisms evaluated in our experiments—LLM-Auction, the first-price variant of RAG-Auction, and the first-price variant of MOSAIC—satisfy monotonicity, all three are IC under this model, and advertisers will bid truthfully. This further supports the validity of our evaluation protocol.

\paragraph{RAG-Auction~\citep{rag_auction}:} RAG-Auction is a pre-generation mechanism that allocates ads by retrieval relevance before response generation. Following the LLM-Auction setting, we use the multi-allocation segment auction configuration and set $K=3$, matching the average insertion volume of LLM-Auction on the test set. We follow the original pseudocode for allocation: the retrieval relevance score $q_i$ is the cosine similarity between Qwen3-Embedding-0.6B embeddings of the ad profile $a_i$ and query $x$. The original RAG-Auction uses second-price payments; under our shared protocol, we keep its allocation rule and recompute realized revenue and reward with first-price click payments.

\paragraph{MOSAIC~\citep{MOSAIC}:} MOSAIC is a post-generation mechanism that samples multiple LLM responses at inference time and selects among them via auction. We reproduce its allocation procedure within our simulation environment and use the same reward model as LLM-Auction, defined in Eq.~\ref{eq:reward_imp}, to ensure a common evaluation signal. Because MOSAIC's original payment rule targets a different serving interface, we report first-price realized payments for the shared allocation-efficiency comparison.

\paragraph{LLM-Auction (Oracle):} In our simulation environment, we can directly obtain feedback from User-LLM for different sampled responses to the same user query, enabling direct reward calculation for DPO-based post-training optimization of the LLM. This baseline effectively provides an approximation of the ideal performance metric prediction over the entire sample space, thus serving as an upper bound reference for LLM-Auction's performance.

\subsection{Implementation Details of Ablation Studies}
\label{appendix:ablation}

\paragraph{Variant with reduced sample size.} During each epoch of the IRPO algorithm in LLM-Auction, we randomly sample a subset of the constructed online dataset $\mathcal{D}_{\text{on}}$ while keeping all other components unchanged. LLM-Auction (50\% samples) in Table~\ref{tab:ablation} represents a variant with the reward model using a pCTR model trained on only half of the samples.

\paragraph{Variant without response context.} As described in Appendix~\ref{appendix:pctr}, our pCTR model uses the text embedding of the ad-integrated response generated by the LLM as one of the input features. We construct the LLM-Auction (w/o response) variant by removing this input feature while keeping the rest of the pCTR model architecture and hyperparameters unchanged. As shown in Table~\ref{tab:ablation}, ablating this input feature significantly impacts the training of LLM-Auction, demonstrating the importance of this design for the LLM to learn an effective allocation rule.

\paragraph{Unified GAUC computation.} We use the LLM-Auction allocation rule after 6 epochs of training to sample responses on held-out test queries and collect click feedback from the User-LLM, constructing a unified evaluation dataset. This dataset approximates the data distribution under the optimal allocation, and we compute the GAUC of pCTR models on this evaluation dataset to assess the prediction accuracy of pCTR models corresponding to LLM-Auction under different epochs and settings under identical conditions.

\subsection{Compute Resources}
\label{appendix:compute_resources}

All experiments are conducted on a 32-GPU internal cluster. The GPU compute is approximately equivalent to NVIDIA A100 GPU, and each GPU provides 96 GB memory. Under this setup, one round of LLM alignment training takes approximately two hours.

\section{Additional Experimental Results}
\label{appendix:add_exp_result}

\subsection{Simulator Robustness}
\label{appendix:sim_robustness}

Because the main evaluation uses LLM-simulated click feedback, we test whether the conclusions persist under a different User-LLM. This experiment follows the discussion in Section~\ref{subsec:benchmark}: the simulator provides a controlled sandbox, while LLM-Auction itself only requires click-style feedback and is not tied to the preference distribution of a particular simulator. Table~\ref{tab:user_llm_robustness} replaces Qwen3-4B with Qwen3-14B as the User-LLM, thereby changing the feedback distribution while preserving the same feedback data format.

The absolute revenue and reward values shift under Qwen3-14B, confirming that different User-LLMs induce different click distributions. Nevertheless, LLM-Auction (Oracle) remains far above the pre-trained LLM under both simulators: revenue increases from 70.4 to 223.3 with Qwen3-4B feedback and from 51.6 to 153.0 with Qwen3-14B feedback, while reward changes from negative to positive in both cases. This pattern indicates that LLM-Auction learns from the supplied feedback distribution rather than exploiting specific preference of one User-LLM. In deployment, the same reward-alignment pipeline can replace simulated clicks with real user click logs, so the mechanism can adapt to real-world user click distributions as long as click feedback is available.

\begin{table}[ht]
\caption{Robustness to the User-LLM feedback distribution.}
\label{tab:user_llm_robustness}
\centering
\begin{tabular}{llcc}
\toprule
\textbf{Method} & \textbf{User-LLM} & \textbf{Avg Rev} & \textbf{Avg Rew} \\
\midrule
Pre-trained LLM & Qwen3-4B & 70.4 & -218.0 \\
LLM-Auction (Oracle) & Qwen3-4B & 223.3 & 125.2 \\
Pre-trained LLM & Qwen3-14B & 51.6 & -236.8 \\
LLM-Auction (Oracle) & Qwen3-14B & 153.0 & 90.3 \\
\bottomrule
\end{tabular}
\end{table}

\subsection{Human Validation of the User-LLM Simulator}
\label{appendix:human_validation}

To check whether the User-LLM's click feedback reflects human judgments rather than model-specific artifacts, we conduct a small-scale human validation. We sample 200 ad-integrated responses generated by a diverse set of LLMs, containing 703 ad impressions in total. Three independent in-house professional annotators are shown exactly the same context as the User-LLM (user profile, query, and response) and make a click decision for each inserted ad. All three annotators agree on 95.3\% of the impressions, and disagreements are resolved by majority vote.

For each of the $M=703$ impressions, let $h_j \in \{0,1\}$ denote the human majority label and $\ell_j \in \{0,1\}$ the User-LLM's click decision. We report the agreement rate and Cohen's $\kappa$:
\begin{equation*}
    \text{Agreement}=\frac{1}{M}\sum_{j=1}^{M}\mathbbm{1}[\ell_j=h_j], \qquad \kappa=\frac{p_o-p_e}{1-p_e},
\end{equation*}
where $p_o$ is the observed agreement and $p_e=\hat{P}(\ell=1)\hat{P}(h=1)+\hat{P}(\ell=0)\hat{P}(h=0)$ is the agreement expected by chance from the two marginal click rates. Cohen's $\kappa$ thus discounts chance agreement ($\kappa=0$: chance level; $\kappa=1$: perfect agreement), so it is not inflated by the imbalance between clicked and non-clicked impressions. We run each User-LLM three times and report the mean $\pm$ std over the three runs in Table~\ref{tab:human_validation}.

\begin{table}[ht]
\caption{Agreement between User-LLM click decisions and human majority labels on 703 ad impressions (mean $\pm$ std over 3 runs).}
\label{tab:human_validation}
\centering
\begin{tabular}{lcc}
\toprule
\textbf{User-LLM} & \textbf{Agreement} & \textbf{Cohen's $\kappa$} \\
\midrule
Qwen3-4B (main experiments) & 75.3\% $\pm$ 0.6\% & 0.49 $\pm$ 0.01 \\
Qwen3-30B-A3B & 77.8\% $\pm$ 1.2\% & 0.52 $\pm$ 0.02 \\
Claude Sonnet 4.6 & 79.9\% $\pm$ 1.1\% & 0.54 $\pm$ 0.03 \\
\bottomrule
\end{tabular}
\end{table}

The User-LLM agrees with human judgments well above chance for all three models, and the agreement increases with stronger judge LLMs. This indicates that the simulator's feedback is not merely a model-specific artifact. Further aligning the User-LLM with human behavior through post-training or prompt optimization is a valuable direction for building more realistic user simulation environments.

\subsection{Zero-Shot Generalization}
\label{appendix:zero_shot}

We evaluate zero-shot transfer along two axes: replacing the candidate ad pool and replacing the user population. These two settings test different deployment risks. The ad-side experiment asks whether LLM-Auction can handle new advertiser domains without additional training, while the user-side experiment asks whether the learned allocation behavior transfers to heterogeneous user-profile distributions.

\begin{table}[ht]
\caption{Zero-shot generalization across the ad dimension. The candidate advertisement pool is replaced with 100 previously unseen ads, and the model is evaluated without additional training.}
\label{tab:zero_shot_ads}
\centering
\begin{tabular}{lcc}
\toprule
\textbf{Method} & \textbf{Avg Rev} & \textbf{Avg Rew} \\
\midrule
Pre-trained LLM & 66.2 & -231.0 \\
LLM-Auction (Oracle) & 169.7 & 56.8 \\
\bottomrule
\end{tabular}
\end{table}

\noindent\textbf{Generalization across ads.}
Table~\ref{tab:zero_shot_ads} replaces the entire candidate advertisement pool with 100 previously unseen ads. LLM-Auction (Oracle) is evaluated directly, without additional training, and still improves revenue from 66.2 to 169.7 and reward from $-231.0$ to 56.8. This result suggests that the mechanism does not simply memorize training advertisements. Instead, it transfers allocation behavior based on generalizable signals such as query-ad semantic relevance, response context, and bids, which are precisely the signals available when new advertisers enter a platform.

\begin{table}[ht]
\caption{Zero-shot generalization across the user dimension. The evaluation uses 100 unseen user profiles and 500 newly generated queries, with no additional training.}
\label{tab:zero_shot_users}
\centering
\begin{tabular}{lcc}
\toprule
\textbf{Method} & \textbf{Avg Rev} & \textbf{Avg Rew} \\
\midrule
Pre-trained LLM & 79.1 & -219.3 \\
LLM-Auction (Oracle) & 242.6 & 136.8 \\
\bottomrule
\end{tabular}
\end{table}

\noindent\textbf{Generalization across users.}
Table~\ref{tab:zero_shot_users} evaluates 100 unseen user profiles that generate 500 new queries. LLM-Auction (Oracle) again retains large gains, increasing revenue from 79.1 to 242.6 and reward from $-219.3$ to 136.8. This user-side transfer indicates that the learned policy is not tied to the exact synthetic profiles or query set used during training. Instead, it generalizes to new user interests and information needs, supporting the applicability of LLM-Auction to heterogeneous user populations.

\subsection{Base Model Scaling}
\label{appendix:model_scaling}

LLM-Auction integrates auction decision-making and ad-integrated response generation into a single Ad-LLM. Thus, the base model controls both the language-generation quality and the capacity to internalize allocation signals. Table~\ref{tab:model_scaling} compares Qwen3-4B with Qwen3-30B-A3B to examine how this choice affects performance.

Scaling the base model improves both the pre-trained LLM baseline and LLM-Auction after LoRA tuning. The pre-trained Qwen3-30B-A3B baseline already outperforms Qwen3-4B, suggesting that stronger language models provide better initial response and ad-selection behavior. More importantly, LLM-Auction (Oracle, LoRA) improves from 138.2 to 236.9 in revenue and from 25.4 to 103.6 in reward when the base model is scaled up. These results show that LLM-Auction inherits the scaling properties of the underlying generator: stronger models are better able to use feedback, context, and bid signals during alignment. We also observed training instability and collapse with smaller Qwen3-1.7B models, indicating that sufficient base-model capability is important for reliable auction-oriented alignment.

\begin{table}[ht]
\caption{Effect of base-model choice on allocation performance.}
\label{tab:model_scaling}
\centering
\begin{tabular}{llcc}
\toprule
\textbf{Method} & \textbf{Base model} & \textbf{Avg Rev} & \textbf{Avg Rew} \\
\midrule
Pre-trained LLM & Qwen3-4B & 70.4 & -218.0 \\
Pre-trained LLM & Qwen3-30B-A3B & 101.4 & -135.7 \\
LLM-Auction (Oracle, LoRA) & Qwen3-4B & 138.2 & 25.4 \\
LLM-Auction (Oracle, LoRA) & Qwen3-30B-A3B & 236.9 & 103.6 \\
\bottomrule
\end{tabular}
\end{table}

\section{Qualitative Analysis}
\label{appendix:qualitative}

We present qualitative comparisons of ad-integrated responses to illustrate the performance improvements achieved by our proposed LLM-Auction mechanism. We compare responses generated by the pre-trained LLM  (before LLM-Auction training) and the fine-tuned model of LLM-Auction, given identical user queries, candidate ads, and bids. 
Note that our experiments use a relatively small-scale LLM (Qwen3-4B), and the design of the reward model focuses on mechanism objectives rather than the quality of the response. As a result, some responses generated may contain artifacts related to the reasoning process for the insertion of the ad.
For presentation clarity, such extraneous content has been removed from the examples below.

\paragraph{Better query-ad relevance.}
Table~\ref{tab:quality_exp1} demonstrates improved semantic matching between user queries and selected advertisements. The pre-trained LLM inappropriately inserts a sofa advertisement (Ad-9420) in response to a query about preparing nourishing soups for elderly and children, exhibiting poor query-ad alignment. In contrast, LLM-Auction selects a vitamin B12 supplement advertisement (Ad-6542), which is directly related to the nutritional and wellness context of the query. Although this selection is constrained by our limited ad pool, it demonstrates better thematic relevance. This improvement indicates that LLM-Auction successfully learns to evaluate and prioritize ads based on semantic relevance with user queries, a critical factor for user engagement and click-through probability.

\paragraph{Improved contextual integration.}
Table~\ref{tab:quality_exp2} illustrates improved contextual integration of advertisements within response narratives. The pre-trained LLM exhibits a tendency to append ads as isolated afterthoughts disconnected from the main content, resulting in non-native placements that disrupt reading flow. By contrast, LLM-Auction seamlessly weaves advertisements into the natural discourse structure. For example, when discussing wetland hiking routes, it organically recommends anti-slip sports shoes (Ad-4064) as practical gear; when describing forest camping experiences, it naturally mentions clip-on Bluetooth earphones (Ad-7444) to enhance the outdoor experience. These contextually-grounded placements create a more cohesive narrative where ads serve as relevant recommendations rather than intrusive interruptions. This native integration style is more likely to be perceived as helpful rather than disruptive, thereby improving user experience and increasing click propensity.

\paragraph{Stronger bid-awareness in generation.}
Table~\ref{tab:quality_exp3} reveals LLM-Auction's acquisition of bid-awareness, a fundamental mechanism property. The pre-trained LLM demonstrates negligible sensitivity to bid signals: when Ad-2 (American Ginseng) submits a substantially higher bid (41) compared to Ad-1 (Progesterone Tea, bid 1), the model fails to adjust allocation accordingly. Moreover, when Ad-1's bid reaches the maximum value (100) while Ad-2 maintains bid 41, the pre-trained model simultaneously displays both advertisements with comparable prominence, effectively diluting Ad-1's competitive advantage despite its willingness to pay higher prices. In contrast, LLM-Auction exhibits clear bid-aware behavior: it increases the prominence and favorable positioning of higher-bidding advertisers in generated responses. This bid-awareness validates that our framework successfully trains the LLM to internalize auction dynamics, ensuring that advertisers receive allocation results corresponding to their submitted bids—a critical property for encouraging advertisers to participate and mechanism credibility in real-world deployment.

\section{Extended Limitations and Future Work}
\label{appendix:extended_limitations}

\paragraph{Benchmark scale and real-world generalization.}
The experiments rely on an LLM-based simulator because public, large-scale real-world logs for LLM-native advertising are not yet available. This simulator provides a controlled sandbox for comparing mechanisms, but it cannot fully capture the diversity of real user intent, long-term satisfaction, trust dynamics, advertiser strategy, or platform constraints. The robustness and zero-shot analyses in Appendix~\ref{appendix:add_exp_result} reduce the risk that the results depend on a single simulator distribution, but production validation should ultimately use real impressions, clicks, conversions, and user-satisfaction signals. A central future direction is therefore to build standardized industrial-scale benchmarks for LLM-native advertising, with richer user populations, broader ad domains, and more faithful user-behavior simulation. Agent-based simulation techniques~\citep{recoworld,user_behavior_simulation_agents} may provide a useful path toward more realistic mechanism evaluation before real deployment logs become available.

\paragraph{Reward design, hallucination, and bias.}
The current reward emphasizes allocation efficiency while using lightweight penalties for excessive ad insertion and formatting errors. This design helps isolate the mechanism objective and already reduces visible user-experience failures, but it remains too coarse for deployment settings where LLM-native ads may affect factuality, credibility, bias, and user trust. For example, an inserted ad may contain or encourage misleading claims, make the surrounding answer less faithful to the user's intent, amplify demographic or topical bias, or blur the boundary between helpful recommendation and sponsored persuasion. LLM-Auction is compatible with richer user-experience signals through the same reward interface: future reward models should incorporate ad fidelity, response factuality, credibility, hallucination risk, bias, disclosure quality, and long-term trust metrics, together with calibrated pCTR models, safety filters, and human or automated audits. These additions would move the framework beyond controlling the number and format of inserted ads toward optimizing whether ad-integrated responses remain trustworthy and aligned with user welfare.

\paragraph{Optimization algorithms and online learning.}
Our current optimization procedure is a multi-stage offline prototype. IRPO periodically refreshes the pCTR model and the Ad-LLM, which allows the allocation policy to track the reward model induced by updated feedback, but it does not yet provide an end-to-end online learning system with strict latency, exploration, rollback, and safety guarantees. The alternating pipeline may also be sensitive to reward-model errors, delayed feedback, distribution shift, and short-term click over-optimization when deployed in a nonstationary market. Future work should treat LLM-Auction as a general mechanism-learning framework that can be paired with stronger optimization algorithms, including advanced reinforcement learning and preference-optimization methods~\citep{ppo,grpo,yu2025dapo,gspo}. Such extensions could use real-time user feedback, conservative policy updates, and safety-constrained exploration to improve online adaptation while preserving response quality and mechanism stability.

\paragraph{Mechanism design beyond first-price payments.}
The first-price payment rule is computationally simple and has favorable incentive properties under the UM and ROI-constrained VM advertiser models analyzed in this paper. It is nevertheless only one point in a larger design space. Real advertising markets include budget pacing, repeated interactions, multi-objective advertisers, strategic exploration, and heterogeneous risk preferences. Designing computationally feasible incentive-compatible payment rules for these broader behaviors remains an important step toward optimal LLM-native advertising mechanisms. This direction also raises new equilibrium questions because the allocation object is a generated response rather than a fixed slot.

\paragraph{Scalability to large ad inventories.}
Our experiments place all candidate ads in the prompt, which does not scale to inventories with thousands of ads. LLM-Auction is compatible with the two-stage pipeline common in industrial advertising systems: a lightweight retrieval module first selects a small candidate set from the full inventory, and generative allocation is applied only to the retrieved candidates, so the LLM-side cost scales with the size of the retrieved set rather than the inventory. If retrieval is bid-independent, an advertiser cannot change its retrieval outcome by misreporting its bid, and the incentive properties in Section~\ref{subsec:payment} carry over to the retrieved candidates. If retrieval depends on bids, a bid also affects whether the advertiser enters the candidate set, so incentive compatibility must account for both stages. Such two-stage auctions have been studied for conventional ad auctions~\citep{two_stage_auction_2022,two_stage_auction_2025}, and extending these designs to generative allocation is an important direction.

\paragraph{Measurement beyond clickable ads.}
Clickable ads make attribution precise in our simulation: the User-LLM can identify inserted ads and return structured click feedback. This abstraction is useful for controlled experimentation, but LLM-native advertising may influence users through narrative framing, product mentions, brand recall, and delayed conversions even when no explicit click occurs. Future work should develop content-based measurement methods that better match how users interact with generated answers, including metrics for helpfulness, disclosure, trust, long-term satisfaction, and advertiser value. Such measurement tools will be foundational for mature LLM advertising ecosystems.

\paragraph{Model dependence and deployment readiness.}
The model-scaling experiment shows that stronger base LLMs improve allocation performance, while smaller models can become unstable during training. This dependence is expected because LLM-Auction asks one generator to satisfy language quality, ad relevance, bid awareness, mechanism constraints, and formatting requirements simultaneously. Future work should study parameter-efficient training, distillation, stability regularization, monitoring, and rollback mechanisms so that LLM-native advertising mechanisms can be reliably deployed in high-concurrency industrial environments without requiring the largest available model.

\begin{table}[p]
\caption{A comparison example of ad-integrated response generated by the pre-trained LLM and LLM-Auction. The format of the inserted clickable ads is \texttt{@ad\_title@[ad\_id]}, and we mark the corresponding text with {\color{blue}\underline{blue underlines}}.}
\label{tab:quality_exp1}
\begin{small}
% \fontsize{8pt}
\begin{tabularx}{\textwidth}{@{}X@{}}
\toprule
\textbf{User Query:} \\
How to make nutritious and easy-to-digest health soup with affordable ingredients for the elderly and children in autumn and winter? \\
\midrule
\textbf{Ad-Integrated Response (Pre-trained LLM):} \\
During the autumn and winter seasons, the digestive systems of the elderly and children tend to be more sensitive. Therefore, nourishing soups should be light, easy to digest, and rich in nutrients. Here are some affordable yet nutritious and gentle soup recommendations:
\begin{enumerate}
    \item Pork Rib and Radish Soup: Pork ribs are rich in calcium and protein, while radishes are mild and aid digestion. Combining the two helps replenish qi and blood while soothing the stomach. Simply blanch the ribs, then simmer them with chopped white radish and a few slices of ginger for about an hour. This soup is ideal for both the elderly and children.
    \item Yam and Millet Porridge: Yam strengthens the spleen and stomach, while millet soothes digestion and promotes relaxation. This porridge is highly suitable for both children and seniors. Peel and cube the yam, then cook it with millet until soft. Adding a few red dates enhances the flavor while improving appetite and digestion.
    \item Cabbage and Tofu Soup: Cabbage is neutral in nature and rich in vitamin C, while tofu is rich in plant protein. This combination is simple and nutritious. Simply slice the cabbage, cut the tofu into cubes, add a little salt and ginger slices, and bring to a boil. It's a good choice for children and the elderly to drink daily.
\end{enumerate}
These soups are not only inexpensive and easy to prepare but also meet the nutritional needs of the autumn and winter seasons, making them perfect for daily family meals. For enhanced taste or nutrition, a small amount of goji berries or red dates can be added, though excessive amounts should be avoided to prevent digestive discomfort.
{\color{blue}\underline{@Parsmenn Minimalist Cloud Sofa@[Ad-9420]}} Though unrelated to food, if your family includes elderly or children who enjoy a cozy home environment, this sofa can enhance overall living comfort, making wellness moments even more relaxing. \\
\midrule
\textbf{Ad-Integrated Response (LLM-Auction):} \\
During the autumn and winter seasons, the digestive systems of the elderly and children are relatively weaker. Therefore, nourishing soups should prioritize easy digestion, gentle ingredients, and nutrient density. Below are some affordable yet nutritious and easily digestible soup recommendations—budget-friendly and suitable for those with weak constitutions or sensitive stomachs.
\begin{enumerate}
    \item Yam and Pork Rib Soup: Yam strengthens the spleen, moistens the lungs, and supports kidney function, while lean ribs provide high-quality protein and calcium. This mild and nourishing soup is particularly suitable for the elderly and children. Both yam and ribs are common and inexpensive ingredients, requiring only minimal seasoning.
    \item Lily Bulb, Lotus Seed, and White Fungus Soup: Lily bulbs and lotus seeds help calm the mind, nourish the heart, and relieve coughs, while white fungus is rich in collagen, moisturizing dryness—ideal for the dry autumn and winter months. This oil-free, light soup is gentle on the stomach, promoting better sleep for both children and seniors.
    \item Pumpkin and Millet Porridge: Although not a traditional soup, this porridge made by cooking pumpkin and millet is both nutritious and easy to digest. Pumpkin is rich in vitamin A, which helps boost immunity, while millet nourishes the spleen and stomach, making it an excellent choice for daily health maintenance for children and the elderly.
\end{enumerate}
All of these soups use readily available, low-cost ingredients and are simple to prepare, making them perfect for everyday family meals. If you wish to enhance flavor or nutrition, a small amount of goji berries or red dates can be added, but moderation is key to avoid digestive strain.
{\color{blue}\underline{@Staid Vitamin B12@[Ad-6542]}}—Vitamin B12 helps improve nervous system function, supporting mental clarity and appetite in both the elderly and children. Especially in autumn and winter, when deficiencies are more common, it can serve as a beneficial dietary supplement. \\
\bottomrule
\end{tabularx}
\end{small}
\end{table}

\begin{table}[p]
\caption{A comparison example of ad-integrated response generated by the pre-trained LLM and LLM-Auction. The format of the inserted clickable ads are \texttt{@ad\_title@[ad\_id]}, and we mark the corresponding text with {\color{blue}\underline{blue underlines}}.}
\label{tab:quality_exp2}
\begin{small}
\begin{tabularx}{\textwidth}{@{}X@{}}
\toprule
\textbf{User Query:} \\
What are some safe and nature-rich outdoor camping or hiking routes near Shanghai suitable for families on weekends? \\
\midrule
\textbf{Ad-Integrated Response (Pre-trained LLM):} \\
There are many outdoor camping and hiking routes near Shanghai that are perfect for family weekends, offering both natural scenery and a safe, engaging experience for children. Here are some recommended options:
\begin{enumerate}
    \item Jinshan Coastal Camping Park. Located in Jinshanwei, about an hour's drive from downtown Shanghai, this park features open beaches, wetlands, and forests, making it an ideal spot for family camping. The park offers tent rentals, a children’s activity area, and parent-child workshops, with well-maintained safety facilities and dedicated childcare staff. Kids can enjoy activities like sand sculpting, shell collecting, and birdwatching.
    \item Hiking Trail at Sheshan National Forest Park (Songjiang). With an elevation of around 90 meters, Sheshan has gentle terrain and clearly marked trails, including kid-friendly paths and observation decks suitable for children aged 3 and above. The route is lined with pine trees, maples, and wildflowers, offering fresh air and scenic views, along with rest pavilions and drinking water stations. Parents can teach children about plant identification and insect observation to enhance their connection with nature.
    \item Water Town Hiking Routes Around Zhujiajiao Ancient Town (Qingpu). Combining culture and nature, the area around Zhujiajiao Ancient Town features historic bridges, small rivers, and pavilions, making it great for family biking or walking. You can choose trails along the Xishi River or Liantang River, where kids can learn about the history of Jiangnan water towns while participating in hands-on activities like making sachets or sketching scenic views.
\end{enumerate}
All these routes prioritize safety and interactivity. It’s recommended to book campsites or tickets in advance through official channels or local travel platforms to ensure a secure and enjoyable experience. Additionally, bring essentials like sunscreen, drinking water, and a first-aid kit to ensure a smooth outdoor adventure.

{\color{blue}\underline{@FILA Kids Basketball Shoes@[Ad-7633]}} are a practical choice for family outings, providing comfort and anti-slip designs suitable for wet forest trails or grassy areas. \\
\midrule
\textbf{Ad-Integrated Response (LLM-Auction):} \\
Here are some family-friendly weekend camping and hiking routes around Shanghai that balance safety and nature immersion. These recommendations prioritize accessibility, scenic beauty, and security, making them ideal for family outings:
\begin{enumerate}
    \item Zhujiajiao Wetland Hiking (Qingpu District). The wild wetland park near Zhujiajiao Ancient Town offers gentle hiking trails with clear waters and reed marshes—perfect for children to explore aquatic ecosystems. The routes are set in safe areas outside the town, featuring viewing platforms and non-slip walkways without steep slopes or complex terrain. Parents and kids can collect fallen leaves, observe insects, and learn about nature together. {\color{blue}\underline{@Fila Kids' Sports Shoes@[Ad-4064]}}—For outdoor activities, slip-resistant, comfortable, and durable shoes are essential for children. These Fila kids' shoes feature anti-slip designs suitable for wet surfaces and hiking. Parents should prepare a pair for their little adventurers.
    \item Sheshan National Forest Park Camping Area. With an elevation of about 100 meters, Sheshan boasts well-preserved mountain landscapes and lush vegetation, making it a popular camping spot for Shanghai residents. The camping zone includes a secured children's tent area, fire-safe zones, first aid stations, and a dedicated kids' activity space with nature education guides. Clear weekends are ideal for stargazing, allowing children to appreciate the beauty of the night sky. {\color{blue}\underline{@Sanag Clip-on Bluetooth Earphones@[Ad-7444]}}—During outdoor camping, children may get distracted by wind or noise. These bone-conduction earphones sit comfortably outside the ears, ensuring safety while enhancing their forest experience with clear audio and hearing protection.
    \item Jinshan Langxia Ecological Park Hiking Trail. Located about 40 km (1-hour drive) from downtown Shanghai, this is one of the top-recommended family hiking routes. The park features open forest trails, meandering streams, and abundant native flora and birds. With flat paths, rest areas, and child-friendly facilities, it suits kids aged 3+. Families can enjoy picnics in the woods, fostering both observation skills and physical stamina in a serene natural setting.
\end{enumerate}
A warm and practical insulated water bottle is recommended during camping to avoid digestive discomfort from cold drinks. \\
\bottomrule
\end{tabularx}
\end{small}
\end{table}

\begin{table}[p]
\caption{A comparison example of ad-integrated response generated by the pre-trained LLM and LLM-Auction. The format of the inserted clickable ads are \texttt{@ad\_title@[ad\_id]}, and we mark the corresponding text with {\color{blue}\underline{blue underlines}}.}
\label{tab:quality_exp3}
\begin{small}
\begin{tabularx}{\textwidth}{@{}X@{}}
\toprule
\textbf{User Query:} \\
The weather has turned cooler recently. Could you recommend some herbal teas suitable for middle-aged and elderly people for health maintenance? \\
\midrule
\textbf{Ad-Integrated Response (Pre-trained LLM), Ad-1 (Progesterone Tea) Bid 1, Ad-2 (American Ginseng) Bid 41:} \\
As the weather turns cooler, middle-aged and elderly individuals need to pay extra attention to staying warm and nurturing their health. Moderately consuming warm, nourishing, and moisture-balancing herbal teas, paired with a balanced diet, can help boost immunity and improve overall constitution. Below are some recommended health teas:
\begin{enumerate}
    \item Astragalus and Goji Berry Tea: Astragalus replenishes qi (vital energy), while goji berries nourish the kidneys and brighten the eyes. Ideal for those experiencing fatigue or qi deficiency. Steep 10g of astragalus and 15g of goji berries in boiling water daily; this tea can be consumed long-term.
    \item Angelica and Ginger Tea: Suitable for individuals with cold body constitution or cold hands/feet, as it warms meridians, dispels cold, and promotes circulation. Simmer 6g of angelica root and 3 slices of ginger in water as a daily brew.
\end{enumerate}
For a thoughtful wellness gift, consider the {\color{blue}\underline{@Progesterone Tea@[Ad-2489]}}. This gentle yet potent formula combines traditional herbal elements, making it an excellent choice for daily health maintenance in middle-aged and elderly adults.\\
\midrule
\textbf{Ad-Integrated Response (Pre-trained LLM), Ad-1 (Progesterone Tea) Bid 100, Ad-2 (American Ginseng) Bid 41:} \\
As the weather turns cooler, middle-aged and elderly individuals, with their sensitive constitutions, may benefit from drinking warm, nourishing herbal teas that support digestion and regulate qi and blood. Pairing these with a balanced diet can help boost immunity, improve sleep, and relieve joint discomfort. Here are some recommendations:
\begin{enumerate}
    \item Progesterone Tea: Contains ingredients like flaxseed, figs, mulberries, and black goji berries. It helps regulate female endocrine function, alleviates fatigue, and improves sleep quality, making it ideal for middle-aged and elderly women as a daily beverage. {\color{blue}\underline{@Progesterone Tea@[Ad-2489]}}.
    \item American Ginseng, Astragalus, and Goji Berry Tea: This blend replenishes qi and blood, enhances stamina, and reduces fatigue, making it especially suitable for middle-aged and elderly individuals who experience weakness or tiredness. For a milder flavor, steep with a few red dates or longan. {\color{blue}\underline{@American Ginseng Tea@[Ad-3356]}}.
\end{enumerate}
\\
\midrule
\textbf{Ad-Integrated Response (LLM-Auction), Ad-1 (Progesterone Tea) Bid 1, Ad-2 (American Ginseng) Bid 41:} \\
As the weather turns cooler, middle-aged and elderly individuals, who generally have weaker constitutions, need to pay extra attention to staying warm and maintaining their health. Below are some recommended health-boosting teas and recipes suitable for this age group, designed to enhance immunity, regulate qi and blood, and alleviate fatigue.

Recommended Health Teas:\\
For example, you might try {\color{blue}\underline{@American Ginseng Tea@[Ad-3356]}}. This tea is rich in natural ingredients such as astragalus, American ginseng, and goji berries, which help replenish qi and blood while strengthening the body. It is particularly suitable for consumption during autumn and winter. \\
\midrule
\textbf{Ad-Integrated Response (LLM-Auction), Ad-1 (Progesterone Tea) Bid 100, Ad-2 (American Ginseng) Bid 41:} \\
As the weather turns cooler, it’s the perfect season for health maintenance. Middle-aged and elderly individuals can benefit from warm, nourishing teas that moisturize dryness and regulate qi and blood. Paired with a balanced diet, these teas may help boost immunity, improve sleep, and ease joint discomfort.

{\color{blue}\underline{@Progesterone Tea@[Ad-2489]}} is a wellness tea blend specially designed for middle-aged and senior women. With natural ingredients like flaxseeds, mulberries, and black goji berries, it supports endocrine regulation and enhances blood circulation—ideal for those experiencing cold hands/feet or frequent late nights. \\
\bottomrule
\end{tabularx}
\end{small}
\end{table}

%%%%%%%%%%%%%%%%%%%%%%%%%%%%%%%%%%%%%%%%%%%%%%%%%%%%%%%%%%%%

\newpage
\section*{NeurIPS Paper Checklist}

\begin{enumerate}

\item {\bf Claims}
    \item[] Question: Do the main claims made in the abstract and introduction accurately reflect the paper's contributions and scope?
    \item[] Answer: \answerYes{}
    \item[] Justification: The abstract and Section~\ref{sec:intro} state the scope, contributions, and experimental claims of LLM-Auction. Sections~\ref{sec:auction_gpt} and~\ref{sec:experiments}, together with the limitations in Section~\ref{sec:limitations} and Appendix~\ref{appendix:extended_limitations}, support these claims and bound their scope.

\item {\bf Limitations}
    \item[] Question: Does the paper discuss the limitations of the work performed by the authors?
    \item[] Answer: \answerYes{}
    \item[] Justification: Section~\ref{sec:limitations} and Appendix~\ref{appendix:extended_limitations} discuss the limitations in detail.

\item {\bf Theory assumptions and proofs}
    \item[] Question: For each theoretical result, does the paper provide the full set of assumptions and a complete (and correct) proof?
    \item[] Answer: \answerYes{}
    \item[] Justification: The assumptions and theoretical statements are given in Sections~\ref{pre} and~\ref{sec:auction_gpt}, including Assumptions~\ref{def:unbiased pctr} and~\ref{assump:bid_independence}. Full proofs of Theorem~\ref{thm:alloc-mono}, Theorem~\ref{thm:continuity}, Theorem~\ref{thm:um_equilibrium}, and Theorem~\ref{thm:vm_roi}, as well as the closed form of the DSIC payment, are provided in Appendix~\ref{appendix:proofs}.

    \item {\bf Experimental result reproducibility}
    \item[] Question: Does the paper fully disclose all the information needed to reproduce the main experimental results of the paper to the extent that it affects the main claims and/or conclusions of the paper (regardless of whether the code and data are provided or not)?
    \item[] Answer: \answerYes{}
    \item[] Justification: Sections~\ref{subsec:benchmark}--\ref{subsec:diagnostics} describe the simulation environment, dataset construction, baselines, metrics, and evaluation protocol. Appendices~\ref{appendix:env} and~\ref{appendix:exp_details} provide prompt templates, model configurations, reward design, pCTR architecture, training procedure, and baseline implementation details.

\item {\bf Open access to data and code}
    \item[] Question: Does the paper provide open access to the data and code, with sufficient instructions to faithfully reproduce the main experimental results, as described in supplemental material?
    \item[] Answer: \answerYes{}
    \item[] Justification: The data and code will be open-sourced in a GitHub repository after completing internal approval.

\item {\bf Experimental setting/details}
    \item[] Question: Does the paper specify all the training and test details (e.g., data splits, hyperparameters, how they were chosen, type of optimizer) necessary to understand the results?
    \item[] Answer: \answerYes{}
    \item[] Justification: Section~\ref{sec:experiments} and Appendix presents all the details.

\item {\bf Experiment statistical significance}
    \item[] Question: Does the paper report error bars suitably and correctly defined or other appropriate information about the statistical significance of the experiments?
    \item[] Answer: \answerYes{}
    \item[] Justification: Table~\ref{tab:main_results} reports the mean $\pm$ standard deviation of all main metrics over 10 disjoint repetition groups of sampled responses (3 independent full runs for MOSAIC), which captures response-sampling variability. Table~\ref{tab:human_validation} reports the mean $\pm$ standard deviation over 3 User-LLM runs.

\item {\bf Experiments compute resources}
    \item[] Question: For each experiment, does the paper provide sufficient information on the computer resources (type of compute workers, memory, time of execution) needed to reproduce the experiments?
    \item[] Answer: \answerYes{}
    \item[] Justification: Refer to Appendix~\ref{appendix:compute_resources}.
    
\item {\bf Code of ethics}
    \item[] Question: Does the research conducted in the paper conform, in every respect, with the NeurIPS Code of Ethics \url{https://neurips.cc/public/EthicsGuidelines}?
    \item[] Answer: \answerYes{}
    \item[] Justification: The work uses synthetic profiles, LLM-simulated click feedback, and controlled offline evaluation rather than collecting private human-subject data. The paper also discusses deployment risks and mitigation directions for trust, safety, bias, and online optimization in Section~\ref{sec:limitations} and Appendix~\ref{appendix:extended_limitations}.

\item {\bf Broader impacts}
    \item[] Question: Does the paper discuss both potential positive societal impacts and negative societal impacts of the work performed?
    \item[] Answer: \answerYes{}
    \item[] Justification: The paper does not include a standalone broader-impact section, but it briefly discusses relevant positive and negative impacts in Section~\ref{sec:intro}, Section~\ref{sec:limitations}, and Appendix~\ref{appendix:extended_limitations}.
    
\item {\bf Safeguards}
    \item[] Question: Does the paper describe safeguards that have been put in place for responsible release of data or models that have a high risk for misuse (e.g., pre-trained language models, image generators, or scraped datasets)?
    \item[] Answer: \answerNA{}
    \item[] Justification: The submission does not release a high-risk pre-trained model, image generator, or scraped dataset. Appendix~\ref{appendix:extended_limitations} discusses future safeguards such as safety filters, audits, monitoring, rollback, and richer reward signals for deployment.

\item {\bf Licenses for existing assets}
    \item[] Question: Are the creators or original owners of assets (e.g., code, data, models), used in the paper, properly credited and are the license and terms of use explicitly mentioned and properly respected?
    \item[] Answer: \answerYes{}
    \item[] Justification:  All the references are made.

\item {\bf New assets}
    \item[] Question: Are new assets introduced in the paper well documented and is the documentation provided alongside the assets?
    \item[] Answer: \answerNA{}
    \item[] Justification: The paper constructs a simulation environment for evaluation but does not release a new dataset, code package, or model asset as part of this submission. The environment design, prompts, configurations, and examples are documented in Appendix~\ref{appendix:env}.

\item {\bf Crowdsourcing and research with human subjects}
    \item[] Question: For crowdsourcing experiments and research with human subjects, does the paper include the full text of instructions given to participants and screenshots, if applicable, as well as details about compensation (if any)? 
    \item[] Answer: \answerNA{}
    \item[] Justification: The experiments do not involve crowdsourcing. User queries and click feedback used for training and evaluation are generated by User-LLM simulators, as described in Section~\ref{subsec:benchmark} and Appendix~\ref{appendix:env}. The only human involvement is the small-scale validation in Appendix~\ref{appendix:human_validation}, where three in-house professional annotators made click decisions on 200 synthetic ad-integrated responses given the same context as the User-LLM.

\item {\bf Institutional review board (IRB) approvals or equivalent for research with human subjects}
    \item[] Question: Does the paper describe potential risks incurred by study participants, whether such risks were disclosed to the subjects, and whether Institutional Review Board (IRB) approvals (or an equivalent approval/review based on the requirements of your country or institution) were obtained?
    \item[] Answer: \answerNA{}
    \item[] Justification: The paper does not involve crowdsourcing or human-subject research. The human validation in Appendix~\ref{appendix:human_validation} only asks annotators to label click decisions on synthetic ad-integrated responses and collects no personal data. All other evaluations use synthetic profiles and User-LLM-generated feedback rather than data collected from study participants.

\item {\bf Declaration of LLM usage}
    \item[] Question: Does the paper describe the usage of LLMs if it is an important, original, or non-standard component of the core methods in this research? Note that if the LLM is used only for writing, editing, or formatting purposes and does \emph{not} impact the core methodology, scientific rigor, or originality of the research, declaration is not required.
    %this research? 
    \item[] Answer: \answerYes{}
    \item[] Justification: LLMs are central to both the proposed allocation rule and the simulation environment; specifically, we train an LLM to serve as the allocation rule for the auction mechanism, while simultaneously employing the LLM-as-a-judge approach to construct the simulation evaluation environment. For detailed implementation specifics, please refer to the sections on experimental details within the main text and the appendix.

\end{enumerate}

\end{document}